\documentclass[a4paper,11pt]{article}
\pdfoutput=1 

\usepackage{jheppub} 

\usepackage{xcolor}
\usepackage{comment}

\def\arXiv#1{\href{http://arxiv.org/abs/#1}{arXiv:#1}}
\def\arXiv#1#2{\href{http://arxiv.org/abs/#1}{arXiv:#1}}
\def\arXivid#1#2{\href{http://arxiv.org/abs/#1/#2}{#1/#2}}
\def\doi#1#2{\href{http://doi.org/#1}{#2}}

\title{\boldmath Quantum complexity and bulk timelike singularities}

\author[a]{Gaurav Katoch,}
\author[b]{Jie Ren}
\author[a]{and Shubho R. Roy}

\affiliation[a]{Department of Physics\\ Indian Institute of Technology, Hyderabad\\ Kandi, Sangareddy, Telengana 502285, India}
\affiliation[b]{School of Physics, Sun Yat-sen University, Guangzhou, 510275, China}

\emailAdd{gauravitation@gmail.com}
\emailAdd{renjie7@mail.sysu.edu.cn}
\emailAdd{sroy@phy.iith.ac.in}

\date{\today}

\abstract{Quantum complexity has already shed light on CFT states dual to bulk geometries containing spacelike singularities \cite{Barbon:2015ria, Bolognesi:2018ion, Caputa:2021pad}. In this work, we turn our attention to the quantum complexity of CFT/quantum gravity states which are dual to bulk geometries containing a naked timelike singularity. The appearance of naked timelike singularities in semiclassical gravity is allowed in string theory, particularly in the context of holography, so long as they satisfy the \emph{Gubser criterion} \cite{Gubser:2000nd, Gursoy:2008za} - those naked timelike singularities which arise as the extremal limits of geometries containing cloaked singularities are admissible. In this work, we use holographic complexity as a probe on geometries containing naked timelike singularities and explore potential relation to the Gubser criterion for detecting allowable naked timelike singularities. We study three specific cases of naked timelike singularities, namely the negative mass Schwarzschild-AdS spacetime, the timelike Kasner-AdS \cite{Ren:2016xhb} and Einstein-dilaton system \cite{Ren:2019lgw}. The first two cases are outright ruled out by the Gubser criterion while the third case is more subtle - according to the Gubser criterion the singularity switches from forbidden to admissible as the parameter $\alpha$ is dialed in the range $[0,1]$ across the transition point at $\alpha = 1/\sqrt{3}$. We probe all three geometries using two holographic complexity prescriptions, namely CA and CV. For the case of the negative mass SAdS and timelike Kasner-AdS$_4$ the complexities display no sign of pathology (both receive finite contribution from the naked singularity). For the Einstein-Dilaton case, action-complexity does display a sharp transition from physical positive values to patholgical negative divergent values (arising from the singularity) as one transcends the Gubser bound. Our study suggests that neither action-complexity (CA) nor volume-complexity (CV) can serve as a sensitive tool to investigate (naked) timelike singularities.}

\arxivnumber{2303.02752}

\begin{document}
\maketitle
\flushbottom

\section{Introduction \& Summary}

The issue of the resolution of spacetime singularities in general relativity is one of the biggest outstanding issues in quantum gravity. Spacetime singularities are inevitable end-points of gravitational collapse of matter-energy \cite{Penrose:1964wq, Hawking:1970zqf}.  In such situations, general relativity breaks down and new UV degrees of freedom are believed to take over. It is a general consensus that such new UV physics will resolve the singularities by smoothing them out, e.g. in string theory, the physics of new degrees of freedom such as strings and branes might remove or resolve the singularities arising in semiclassical gravity \cite{McGreevy:2005ci}. Various isolated examples of singularity resolution are known in string theory, eg. \cite{Strominger:1995cz, Dabholkar:2004dq} (see \cite{key} for an overview). However, it is fair to say that so far in string theory (or for that matter in any UV-complete theory of quantum gravity), there is no universal or systematic mechanism for resolving generic spacetime singularities. Spacetime singularities come in three varieties, namely spacelike, timelike and null. Spacelike singularities those where entire space essentially ends or collapses at some given moment of time, e.g. the big bang (crunch) cosmological singularity or at the heart of neutral black holes. Since all of space collapses, there is no way to evade or avoid this crunch. Timelike singularities, on the other hand, are localized in some compact spatial region and one can in principle stay away from it at all times. Being timelike, some timelike singularities extend all the way up to past timelike infinity and thus constitute a singularity in the initial (metric) data itself! No wonder many researchers perhaps regard timelike singularities as representing unphysical or pathological configurations which should not be part of any UV complete theory of gravity. However, there are no rigorous results regarding the necessary and/or sufficient conditions in string theory which govern the resolution of generic timelike singularities. Part of the reason why generic spacelike singularities, especially cosmological singularities have been intractable in traditional worldsheet sigma model based string theory approaches is due to the fact that these backgrounds are time-dependent and they explicitly break supersymmetry thereby lacking analytical control. See \cite{Liu:2002ft, Liu:2002kb, Cornalba:2002fi, Cornalba:2002nv, Cornalba:2003kd} for some attempts towards this direction. Some have argued in favour of replacing the cosmological singularity by a closed string tachyon condensate \cite{McGreevy:2005ci, Silverstein:2006tm} building on Sen's idea \cite{Sen:2002nu} of the rolling open string tachyon on an unstable brane. Finally nonperturbative setups such as Matrix Models \cite{Craps:2005wd} and the AdS/CFT correspondence have been applied to treat cosmological singularities \cite{Das:2006dz, Das:2006pw, Awad:2008jf} but with modest success. There has been some efforts in dealing with the resolution cosmological singularities in the higher spin gravity set up in as well, see e.g. \cite{Castro:2011fm, Krishnan:2013tza, Burrington:2013dda} in the context of $2+1$-dimensions where the singularity turns out to be a gauge artifact - by performing a higher spin (spin $3$) gauge transformation the metric becomes regular! However, in some respects this is unsatisfactory as turning on higher spin gauge fields can perhaps have more dramatic consequences for observers.
 
The present work has origins in some previous work \cite{Barbon:2015ria, Bolognesi:2018ion} where it was shown that the notion of complexity adds new ways aiding in the investigation/ interpretation or resolution of a class of cosmological singularities like  AdS-Kasner singularity, topological crunch singularity and de Sitter crunch singularity\footnote{Also see more recent work \cite{Jorstad:2023kmq} on probes of black hole singularities using alternative proposals of holographic complexity.}. Holographic complexity is a promising candidate towards providing essential insight in capturing the physics of the singularities. Most transparently, in the action complexity formulation, where the WDW patch receives a direct contribution from the singularity, seems to be an appropriate tool to probe the singularity. In fact, despite appearances, the action and volume complexity seem to agree, this renders complexity in general a viable tool in the investigation of bulk containing the singularities. Those studies clearly suggested that the complexity exhibits a monotonic decrease as one approaches the singularity. This monotonic decrease points to the fact that CFT quantum states have low quantum entanglement at the singularity. This phenomenon is reminiscent of the firewall phenomenon, where the disentangling of gravity degrees of freedom across a black hole horizon leads to the appearance of a naked singularity dubbed as the \emph{firewall} \cite{ Almheiri:2012rt}.
Motivated by the success of this previous work \cite{Barbon:2015ria, Bolognesi:2018ion}, in this work we turn our attention towards the study of naked \emph{timelike}  singularities. Naked timelike singularities are rife in bottom-up holography, in particular in metrics obtained as solutions to the effective holographic theories at zero temperature. Such singularities can be expected, at times, to be resolved by lifting them to full ten-dimensional SUGRA or by the inclusion of string/D-brane states\footnote{There are some isolated examples of the resolution of some innocuous timelike singularities in string theory e.g. the \emph{enhancon} \cite{Johnson:1999qt} where a singular D-brane geometry is resolved when on zooms in close to the singularity and one finds that the D-branes form a shell with flat metric in the interior \cite{Yamaguchi:2001yd}.}. The important question that concerns us is whether a given naked timelike singularity in semiclassical gravity is resolvable in a UV-complete quantum gravity theory e.g. string theory. The chief criterion to answer that issue in the literature is called Gubser criterion \cite{Gubser:2000nd, Gursoy:2008za}. Gubser criterion implies that
 
\emph{Naked singularities arising in bottom-up holography (effective holographic setups) that can be obtained as the extremal deformations of regular blackholes are resolvable in full ten-dimensional (UV complete) string theory.}

 In the following sections, we intend to explore the complexity of the class of spacetimes which comprises of naked timelike singularities. We will first study the simplest negative mass Schwarzschild-AdS spacetime and draw important conclusions regarding the behavior of the complexity. Later we will go ahead and study more complicated examples comprising of the timelike singularities in Kasner and Einstein-Dilaton system.
 
 In AdS/CFT, the dual CFT picture that emerges of the eternal AdS black hole is that of a entangled state of the two copies of CFT living on the asymptotic regions called the thermofield double state \cite{Maldacena:2001kr}. Two such boundaries are joined by an Einstein-Rosen bridge in the bulk spacetime. This ER bridge in the bulk continues to grow long after the boundary field theory attains thermal equilibrium. The spirit of AdS/CFT correspondence begs an answer to the natural question of what dual quantity would suffice to capture this late-time growth.

Susskind conjectured two geometrical duals to address this question and are subsequently called the Complexity Volume \cite{Susskind:2014rva} and the Complexity Action \cite{Brown:2015bva, Brown:2015lvg} conjectures hereafter paraphrased by CV and CA conjecture respectively.

CV conjecture tries to quantify the difficulty in sending a signal across ERB. It proposes that the complexity of the field theory is given by the volume, $V$ of the maximal spacial slice extending into the bulk and terminating on the boundary at the spacial slice on which the quantum state resides. Quantitatively,
\begin{equation}\label{CV}
    C_V(T) =\frac{V_{max}(T)}{G_N L}~,
\end{equation}
where $V(t)$ is the maximal volume of the spacelike slice anchored at the boundary time, $T$. And $L$ is some characteristic length scale associated with the spacetime bulk like AdS radius or horizon radius. However, the choice of this background dependent quantity is ambiguous.

The CA conjecture \cite{Brown:2015bva, Brown:2015lvg} quantifies the holographic dual to the quantum complexity by evaluating the classical bulk action on the Wheeler-DeWitt patch (Wheeler-deWitt patch is the union of all the spacelike slices which extend into the bulk and terminate on the same given spacial slice on the boundary).
Action complexity conjecture posits that, the complexity of the boundary state at time, $T$ is given by
\begin{equation}\label{CA}
C (T) = \frac{I_{WdW}(T)}{\pi \hbar}~,
\end{equation}
where $I_{WdW}$ is the bulk action evaluated on the Wheeler de Witt patch (domain of dependence) for a timelike slice at a time, $t$. Since there is no matter in the bulk, the action is given by \cite{Reynolds:2016rvl,Carmi:2016wjl} the gravitational part
\begin{align*}
I_{WdW}  =  &\frac{1}{16 \pi G_N}\int_{WdW} d^{d+1}x\sqrt{g} \, \left(R-2\lambda\right)  + \, \frac{1}{8 \pi G_N} \int_{\partial WdW} d^dx\sqrt{h} \, K \nonumber\\
&-\frac{1}{8\pi G_N}\int_{\Sigma}d\lambda d^{d-1}\theta\sqrt{\gamma}\kappa+\frac{1}{8\pi G_N}\int_{\Sigma}d^{d-1}x\sqrt{\sigma}a~.
\end{align*}
The first term comprise of the bulk Einstein-Hilbert action term with the cosmological constant, and for the rest of this work, the Gibbons-Hawking boundary term will be evaluated at the timelike boundaries. The third term is the boundary term for the null boundaries of the WdW patch and the constant $\kappa$ comes from writing out the null geodesic equation for the outward directed normal for the null surface, $k$
\begin{align*}
    k^{\mu}\nabla_{\mu}k^{\nu}=\kappa k^{\nu}~.
\end{align*}
There are also the joint contributions which are codimension-two surfaces formed by the intersection of the null-null or null-timelike surfaces. The null boundary contributions are in general complicated and we have just written them out for the sake of completeness.

There some operational advantages in choosing action complexity over volume complexity, the foremost is that, unlike volume complexity, action complexity does not depend upon quantities like arbitrary length scales. The second advantage of action complexity of is that solving for volume complexity is generally a hard variational problem which requires maximization unlike the action complexity wherein one is only required to evaluate the integrals. On the other hand, in geometries which have lesser symmetry, constructing the WdW patch is in itself a nontrivial exercise but the maximal volume slices are relatively easier to construct \cite{Katoch:2022hdf, Bhattacharyya:2022ren}.

The plan of the paper is as follows. In Sec.$\,$\ref{NSAdS} we study the negative mass Schwarzschild AdS geometry. This is known to have a pathological boundary dual, namely a CFT with no stable ground state. This geometry evidently violates the Gubser criterion since it cannot be realized as the extremal (zero temperature) limit of the positive mass Schwarzschild-AdS (SAdS) black hole or for that matter any asymptotically AdS geometry with a cloaked singularity. We first compute the action complexity numerical for a range of the parameters namely the mass parameter, the bulk IR cut off. The action and volume complexity are found to be positive in all cases and display the UV divergences characteristic of a local dual field theory. Notably the GHY term contribution from the singularity vanishes! However the volume complexity is found to lower than that of empty (global) AdS! We speculate this observation to be a potential sign of pathology - anything having a holographic complexity lower than empty AdS perhaps implies instability or some pathology of the dual CFT/ quantum gravity theory. Next, in Sec.$\,$\ref{TKAdS}, we look at an anisotropic asymptotically AdS geometry which is the timelike counterpart of AdS-Kanser spacetime. This contains a naked timelike singularity which too is not allowed according to the Gubser criterion - it cannot be obtained as the extermal limit of a cloaked singularity and instead is obtained via a \emph{Wick rotation} of the more familiar spacelike Kasner-AdS spacetime. We compute the action complexity analytically for this case, and just like in the negative mass Schwarzschild AdS case, it turns out to be perfect physical with the correct type of divergences. The contribution from the singularity this time however is finite and negative definite. Next we work out the volume complexity of this timelike Kasner-AdS geometry and we find that for the range of Kasner exponent, $\alpha<2/3$, the volume complexity is lower than that of empty AdS. Thus according to our speculated complexity criterion, this solution should not be admissible or realizable as semiclassical geometry in any UV complete theory of quantum gravity.  However for the complimentary range of values of the Kasner exponent, $2/3<\alpha<1$, the volume complexity is larger than empty Poincar\'e AdS, and thus appears to be an allowable singularity! So the result as far as volume complexity is concerned is at best mixed. In this second example too we cannot consider either action or volume complexity to be a reliable diagnostic for physically allowable singularities in the semiclassical limit of a QG theory. In the final case, we explore naked singularities arising in the (asymptotically AdS) Einstein-Scalar system in Sec.$\,$\ref{ED}. These solutions are characterized by two parameters, $Q,\alpha$. While $Q$ can be any positive number, $\alpha$ lies in the restricted range $(0,1)$. We show that according to the Gubser criterion when $\alpha\geq \frac{1}{\sqrt{3}}$, the geometry contains an admissible naked singularity while for the complimentary range $\alpha<\frac{1}{\sqrt{3}}$ the singularity is not admissible anymore. Following this we compute the action complexity (both analytically and numerically), and find that for $\alpha<\frac{1}{\sqrt{3}}$ the action contains a negative divergent contribution arising from the singularity, while for $\alpha\geq \frac{1}{\sqrt{3}}$, the complexity is positive and larger than that of empty (global) AdS$_4$. Thus action complexity criterion is in perfect agreement with the Gubser criterion on the appearance of naked timelike singularities in a UV complete quantum gravity theory. We compute the volume complexity next and find that, unlike action complexity, it does not show a sharp transition from negative to positive as the parameter $\alpha$ is dialed from $0$ to $1$. Thus, once again volume complexity fails to discriminate between forbidden and admissible naked timelike singularities. We conclude our work in Sec.$\,$\ref{cc} by discussing our results and provide an outlook for the future investigations regarding the correlation of gravitational singularities with quantum complexity of the holographic dual field theory.

\section{Negative mass Schwarzschild-AdS singularity} \label{NSAdS}
Before delving into a case study of the timelike singularities appearing in the effective holographic theories, we would first like to try out a warm-up example of a timelike singularity, the field theory dual to which is known to be pathological. The negative mass Schwarzschild-AdS geometry is a vacuum solution to the Einstein field equations with a negative cosmological constant, $\lambda = -\frac{\left(D-1\right)\left(D-2\right)}{2\,l^2}$. It is described by the metric (in Schwarzschild coordinates)
\begin{align}\label{SAdS}
    ds^2=-f(r)dt^2+\frac{dr^2}{f(r)}+r^2d\Omega^2_{D-2}~,
\end{align}
where the redshift factor,
\begin{equation*}
f(r)= 1+\frac{r^2}{l^2}+\frac{\mu}{r^{D-3}}
\end{equation*}
has an opposite sign mass term compared to the usual Schwarzschild geometry. Consequently, there is no coordinate singularity at hypersurface $r=2M$, and nor is the boundary of a trapped surface horizon, i.e. an event horizon. However, the hypersurface $r=0$, which is now a \emph{timelike} hypersurface, is still a curvature singularity, and is naked.  In other words, the bulk geometry constitutes of a naked timelike singularity.  The boundary CFT dual to this geometry is pathological, in the sense that it does not admit any stable ground state ($M=0$ is pure AdS which is the vacuum for $M>0$ SAdS geometries)  \cite{Horowitz:1995ta}. We recap the argument which leads to this conclusion in brief here. Let's assume, for the time being, that the negative mass Schwarzschild-AdS geometry is realized in a holographic set up, i.e. it represents some state in an unitary dual CFT. Then, since in holography the ADM mass corresponds to CFT energy, this geometry must be dual to a CFT state of negative energy (below the CFT vacuum state of zero energy). However, since there is no limit to how large the parameter $M$ can be, the CFT must admit states of arbitrary large negative energies. In other words, the dual CFT would not admit a ground state or the CFT Hamiltonian cannot be bounded from below! Thus, we conclude that the negative mass Schwarzschild geometry cannot be realized as a state in healthy CFT (bounded from below). The negative mass SAdS background evidently violates the Gubser criterion - taking an extremal limit of the positive mass Schwarzschild geometry (or for that matter any other charged static asymptotically AdS black hole geometry) does not lead to the negative mass SAdS geometry. Hence according to the Gubser criterion as well, the nakedly singular negative mass SAdS geometry is untenable as a state in string theory or any other UV complete theory of quantum gravity.

We expect the complexity to reflect some pathological behavior that signifies the sickness of the boundary field theory. In the following sections, we work out the complexity using the action complexity conjecture first and later test our claims with the aid of the volume complexity conjecture in the subsequent section.

\subsection{Action complexity for negative mass Schwarzschild AdS}
In this section, we compute action complexity corresponding to the CFT state dual to the negative-mass Schwarzschild AdS black hole. The complexity thus computed is expected to reveal to us the universal divergent pieces. However, on the accounts of the dual CFT lacking any ground state \cite{Horowitz:1995ta}, we expect to register some unphysical characteristics.

\begin{figure}[h]
\begin{center} 
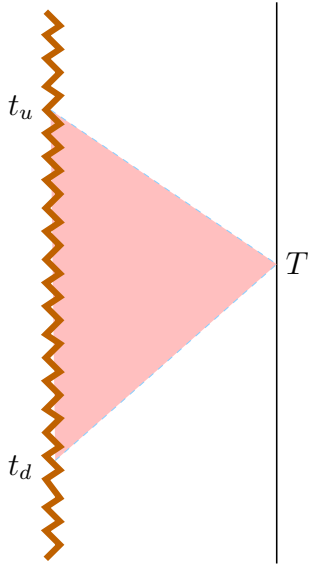
\end{center}
\caption{Penrose diagram for negative mass Schwarzschild-AdS geometry}
	\label{fig6.1}
\end{figure}

Consider a boundary state given at time $t=T$ for which we wish to compute
the complexity as depicted in figure \ref{fig6.1}.
First of all, we need to determine the WdW patch.
For this purpose,  we follow past and future null rays originating at the boundary point $T$. 
The resulting WdW patch is bounded by the null rays 
\[
dt_{u/d}=\mp \frac{dr}{f(r)}~,
\]
 that can readily be written in the integrated form
\[
t_{u/d}(r)=T\pm\int_{r}^{\infty}\frac{dr}{f(x)}~.
\]
The subscript $u$ ($d$) corresponds to the future (past) boundary of the WdW patch. Having determined the boundaries for the WdW patch, we now compute the contributions to the action complexity.
\begin{itemize}
\item \textbf{ Einstein-Hilbert term bulk contribution}
\begin{align}
I_{EH} & =\frac{1}{16\pi G_{N}}\int_{WdW}d^{D}x\;\sqrt{-g}\;(R-2\lambda)~, \nonumber \\
 & =-\frac{(D-1)\:\Omega_{D-2}}{4\pi G_{N}l^{2}}\int_{0}^{\Lambda}dr\:r^{D-2}\:\int_{r}^{\Lambda}\frac{dx}{f(x)}.   \label{SAdS-EH}
\end{align}
The radial integral is IR divergent. So we regulate this contribution by means of the IR regulator, $r\leq\Lambda$. This
contribution is manifestly negative because the integral is positive definite (integrand is positive and the limits of integration are in ascending order).

\item \textbf{The GHY contribution from the timelike singularity}
 On one side, the WdW patch is bounded by the timelike singularity within the range of time coordinates lying between $t_d$ and $t_u$. We will consider a fixed $r$ surface with nonzero $r$ and then set $r\rightarrow0$ at the end. The induced metric on a constant $r$ hypersurface is $$ds^{2}=-f(r)dt^{2}+r^{2}d\Omega_{D-2}^{2}.$$
For which the integral measure can easily be seen to be
\[
d^{D-1}\sqrt{h}=f^{1/2}(r)\:r^{D-2}dt\:d\Omega_{D-2}.
\]
The outward normal to this hypersurface turns out to be
\begin{equation*}
n^{r}=-f^{1/2}(r),\; n^t=n^\Omega=0.
\end{equation*}
The trace of the extrinsic curvature can be found to be equal to $$K=-\frac{f'}{2f^{1/2}}-\frac{D-2}{r}f^{1/2}.$$
The GHY term works out to be
\begin{equation}
I_{GHY}^{r=0} =\lim_{r \rightarrow 0}\;\frac{1}{8\pi G_{N}}\int d^{D-1}x\sqrt{h}\:K=0 \label{SAdS-GHY0},
\end{equation}
where we have considered $D\geq3$.  \emph{Thus the GHY term contribution arising from the singularity vanishes}!

\item  \textbf{The GHY piece from the null boundaries of the WdW patch}\\
\\
 The null boundary of the WdW patch is given by 
\begin{align}
\left( t-T \right)^2 - g(r)^2=0\,\,\, \quad \quad\text{where}\quad
g(r)\equiv\int_{r}^{\infty}\frac{dx}{f(x)}~, \label{SAdS WdW bdry}
\end{align}
According to \cite{Carmi:2016wjl,Reynolds:2016rvl}, the GHY type contribution from the null boundaries is of the following type
\begin{align}  I_{nGHY}=-\frac{1}{8\pi G_N}\int dS \sqrt{\gamma}\,d\lambda \,\kappa. \nonumber
\end{align}
Here $\lambda$ is the parameter parameterizing the null geodesic and $\kappa$ measures the failure of $\lambda$ to be the affine parameter. The area element on the section of constant $\lambda$ on the null boundary is $\sqrt{\gamma} dS $. We choose the null generator on the future null boundary, $\mathcal{N}_+$ of the WdW patch to be $\lambda_{+}=-\frac{r}{l}$, so that the resulting components of the null normal are \begin{align}  
k_{+}^{\mu}&=\left(\frac{l}{f(r)},-l,\Vec{0}\right) 
\end{align}
.Similarly, on the past null surface $\mathcal{N}_-$, we choose $\lambda_{-}=\frac{r}{l}$ to the null parameter so that
\begin{align}
k_{-}^{\mu}&=\left(\frac{l}{f(r)},l,\Vec{0}\right)
\end{align}
This choice of null vectors leads to the affine parametrization $\kappa=0$ and thus the null surface contribution to the action vanishes.

\item \textbf{LMPS counterterm contribution from the null boundary of the WdW patch}
It has been pointed out in \cite{Lehner:2016vdi} that the gravitational action depends upon the choice of the parametrization $\lambda$ on the null boundary components. To restore the coordinate independence upon the gravitational action it is required to include the following counterterms in the action each corresponding to the null boundaries.
\begin{equation}
   I_{LMPS }= -\frac{sign(\mathcal{N})}{8 \pi G_N} \int_{\mathcal{N}} dS \,d\lambda \Theta \log|l \Theta|, \Theta\equiv \frac{1}{\sqrt{\gamma}}\frac{\partial \sqrt{\gamma}}{\partial \lambda}
\end{equation}
where  $\gamma $ is the induced metric on the constant $\lambda$ sections. For the future null boundary, $\mathcal{N}_+$, the constant $\lambda_{+}$ section has the induced metric $ds^2=r^2\,d\Omega_{D-2}^2$ 
\begin{align}
    \Theta_+=-\frac{(D-2)l}{r}
\end{align}
Since $\mathcal{N}_+$ lies to the future of the bulk of WdW patch, $sign(\mathcal{N}_+)=+1$. \\
\begin{align}
I_{LMPS^+}&= -\frac{1}{8 \pi G_N} \int dS \,\int_{-\infty}^{0}d\lambda_+ \Theta_+ \log| \Theta_+| \hspace{3cm} \nonumber \\
&=-\frac{1}{8 \pi G_Nl}\Omega_{D-2}  \,\int_{0}^{\infty} dr\,r^{D-2}\Theta_+ \log|\Theta_+|\nonumber
\end{align}
For the past null boundary $\mathcal{N}_-$ the metric on the constant $\lambda_-$ section is the same as $\mathcal{N}_+$.  But now $sign(\mathcal{N}_-)=-1$ and $\Theta_-=\frac{(D-2)l}{r} = -\Theta_{+}$. So we have,
\begin{align}
    I_{LMPS^-} &= \frac{1}{8 \pi G_N} \int dS \,d\lambda_- \Theta_- \log|\Theta_-|\nonumber\\
    &=-\frac{1}{8 \pi G_N l}\Omega_{D-2}\int_0^{\infty}dr \,r^{D-2}\Theta_+\log|\Theta_+|
\end{align}
Summing up from both the future and past boundaries, 
\begin{align}
    I_{LMPS}&=-\frac{\Omega_{D-2}}{4 \pi G_N l}\int_0^{\Lambda}dr \,r^{D-2}\Theta_+\log|\Theta_+| \nonumber\\
    &= \frac{\Omega_{D-2}}{4 \pi G_N} \left(\frac{\Lambda ^{D-2}}{D-2}-\Lambda ^{D-2} \ln\frac{\Lambda }{(D-2) l}\right) \label{SAdS LMPS}
\end{align}
Evidently this is identical to the LMPS term for pure global AdS as this contribution is independent of $\mu$.

\item \textbf{Joint contributions  from the edges along the intersection of $r=0$ and $\partial WdW$}

The joints are formed by the intersection  of timelike singularity $r=0,$ and null boundaries of WdW patch \eqref{SAdS WdW bdry}. The induced metric  on the joints takes the simple form $$ds^{2}=r^{2}d\Omega_{D-2}^{2}.$$ 
The components of the normals at $r=0$ are
\begin{align*}
    n^{r}&=-f^{1/2}(r)~,&n^{t,\Omega}&=0~.
\end{align*}
And on the null boundaries of the WdW patch, the normal takes the form
\begin{align*}
     k_{\pm}^{r}&= \mp l, & k_{\pm}^{t}&=\frac{l}{f(r)},\, & k_{\pm}^{\Omega} & = 0~.
\end{align*}
Thus, the contribution from each joint is identical and total contribution is
\begin{equation}
    \begin{split}\label{SAdS-joint}
       I_{\mathcal{N}0} & =2\times\frac{1}{8\pi G_{N}}\int_{r=0} d^{D-2}x\:\sqrt{\gamma}\:\ln\left|\frac{n\cdot k_{+}}{l}\right|~,\\
 & =\frac{\Omega_{D-2}}{4\pi G_{N}}\lim_{r\rightarrow0}\left(r^{D-2}\ln\left|\frac{1}{f^{1/2}(r)}\right|\right) = 0. 
    \end{split}
\end{equation}
Thus there are no contributions to the action-complexity from these joints.
 \item \textbf{Null-Null joint contribution: } The null-null joint so formed is at the past (future) end of the null boundary $\mathcal{N}_+$ ($\mathcal{N}_-$) to which the bulk of WdW is also towards the past (future) therefore, the integral receives an overall negative sign. 
 \begin{align}
        \log\left|\frac{k_{+}\cdot k_{-}}{2 l^2}\right|&=-\log \left|f(r)\right|
 \end{align}
 
 The action contribution for the null-null joint is of the form 
 \begin{align}
     I_{nn}&=lim_{r\to \Lambda}-\left( -\frac{1}{8 \pi G_N}\int d^{D-2}\Omega\,r^{D-2} \log \left|f(r)\right|\right) \nonumber \\
     &=\frac{\Omega_{D-2}}{4 \pi G_N}\,\log \left|\frac{\Lambda}{l}\right| \sum_{n=1}^{\left[\frac{D-2}{2}\right]} \frac{(-)^{n-1}}{n} \Lambda^{D-2-2n}\label{SAdS nn}
 \end{align}
This null-null joint contribution is identical to the corresponding contribution in the action-complexity of pure global AdS as evidenced by the fact that it is independent of $\mu$.
 \end{itemize}

The final action complexity for the negative mass Schwarzschild-AdS geometry is given by the sum of the contributions \eqref{SAdS-EH}, \eqref{SAdS-GHY0}, \eqref{SAdS LMPS} and \eqref{SAdS nn}. An exact closed form expression for the action complexity cannot be obtained since the integrals cannot be performed analytically without making any further approximations. One could alternatively choose to evaluate this expression perturbatively, expanding powers of some small parameter such as $\mu/l^{D-3}$ or $\mu/\Lambda^{D-3}$. Instead of making approximations, we evaluate the final action complexity exactly by performing the integrals numerically for a range of values of the mass parameter $\mu$ and the bulk IR cutoff $\Lambda$. For concreteness, the dimensionality of the spacetime has been chosen to be $D=4$ and AdS radius is set to unity, $l=1$. The characteristic features of the action complexity can be surmised from the numerical data provided the table of $C_{A}$ below with the appropriate IR cutoffs and for various values of the mass parameter. 
 \begin{table}[h]
 \centering
 	\begin{tabular}{|l|c|c|r|}
 		\hline
 		{$\mu$} & IR cutoff &$\frac{ G_N \hbar\,}{\Omega_{D-2}}C_A^{SAdS}$ & $\frac{ G_N \hbar \,}{\Omega_{D-2}}(C_A^{SAdS}-C_A^{AdS})$\\
 		\hline
 		2$\times 10^4$ & $10^6$ &$1.75\times 10^{10}$&$0.14$\\
   \hline
 		2$\times 10^5$ & $10^6$ &$1.75\times 10^{10}$&$1.41$\\
 		\hline
 		2$\times 10^6$ & $10^6$ &$1.75\times 10^{10}$& 14.1\\
 		\hline
   	2$\times 10^6$ & $10^8$ &$1.75\times 10^{14}$& 1.0\\
 		\hline
 		2 $\times 10^7$&$ 10^8$ &$1.75\times 10^{14}$&$2.0$\\
 		\hline
 		2$\times 10^8$ &$ 10^8$ &$1.75\times 10^{14}$&$14.5$\\
 		\hline
 	\end{tabular}\\\vspace{0.5cm}
 \caption{Table for dependence upon mass and cut off for negative mass SAdS using CA.}
 \label{table:1}
\end{table}
The following interesting features are evident from the table that the action complexity.
\begin{itemize}
    \item Firstly, the action complexity has strong quadratic dependence upon the bulk IR cutoff. This is easily understood in light of the UV-IR correspondence in the AdS-CFT duality set up. The bulk IR divergences encode the UV divergences of the dual boundary conformal field theory. If the dual boundary theory is a local theory, an extensive quantity such as complexity would naturally be expected to have a leading quadratic divergence (or UV divergence scaling $\Lambda^{D-2}$ for general $D$-dimensional bulk).
     \item Secondly, the action complexity display extremely weak dependence on the negative mass parameter $\mu$, which is not surprising since the dependence on $\mu$ is via the dimensionless combination $\mu/\Lambda^{D-3}$ i.e. is heavily suppressed by powers of the large bulk IR cutoff. The fourth column in the table \ref{table:1} displays the values of complexity after background \textit{(pure-AdS) subtraction}. It is immediately obvious that the difference between the action complexity of negative mass SAdS and the pure AdS geometry is orders of magnitude smaller than the pure AdS complexity and is more significantly negative. 
    \item{Third and perhaps the most couterintuitive observation is that the timelike singularity does not contribute to the complexity at all! In case of the usual (positive mass) Schwarzschild case, the action complexity does receive a nonvanishing and finite (nondiverging) contribution\footnote{In fact this is true for most spacelike singularities, naked or cloaked \cite{Barbon:2015ria, Bolognesi:2018ion}.}}. We interpret this feature to the first sign that complexity is not a tool or probe that is sensitive to the presence of timelike singularties in the bulk.
    \item{Finally complexity expression does not reveal that the dual boundary theory is sick (unstable since does not admit a ground state)}. 
\end{itemize}
According to Gubser criterion \cite{Gubser:2000nd}, since the negative mass Schwarzschild AdS geometry cannot be obtained as the extremal limit of any regular black hole solution, it is an example of unresolvable singular geometry in a fully quantum theory of gravity (string theory) in the sense that it can never be obtained in the semiclassical gravity approximation in a fully quantum theory of gravity. However the action complexity (shown in the third column of table \ref{table:1}) of this unphysical background does not show any obvious signs of such pathology - it is positive definite and scales extensively with dual boundary theory. To complete the analysis we will apply the volume complexity prescription in the next section to compare and contrast with our findings of the action complexity exercise. 
  \subsection{Volume complexity for the negative mass Schwarzschild AdS}
 In this section, we will probe the complexity of negative mass SAdS using the CV prescription. To this end, we consider spacelike hypersurface  $t=t(r)$. The induced metric on this spatial hypersurface given by
 \begin{equation*}
 d\gamma^2=\left(-f(r)t'^2(r
 )+\frac{1}{f(r)}\right)dr^2+r^2d\Omega^2_{D-2}~.
\end{equation*}
We need to extreimize the volume for the codimension-1 spacelike hypersurface anchored at some specific\footnote{Due to the lack of event horizons in this negative mass geometry one has a global timelike killing vector, $\partial_t$, and consequently the complexity is $T$-independent} boundary time, say $T$:
 $$V(T)=\Omega_{D-2}\int^{\infty}_{0}dr\,r^{D-2}\sqrt{-f(r)t'^2(r)+\frac{1}{f(r)}}~.$$
 The variational problem yields the following Euler-Lagrange equation
 \begin{align*}
     2 (D-2) f(r)^3 t'(r)^3-2 f(r) \left((D-2) t'(r)+r t''(r)\right)+r f(r)^2 f'(r) t'(r)^3-3 r f'(r) t'(r)=0~.
 \end{align*}
Assuming the near boundary expansion of the form 
\begin{align*}
    t(r)=T+\frac{t_1}{r}++\frac{t_2}{r^2}+\frac{t_3}{r^3}+\frac{t_4}{r^4}+....~,
\end{align*}
and plugging back into the Euler-Lagrange equation to solve perturbatively term by term yields us the solution $t(r)=T$.
 This solution seems plausible upon exploiting the time translation symmetry to choose the symmetrical solution $t=\text{constant}(=T)$. The maximal volume then turns out to be  \begin{align*}
 V(T)&=\Omega_{D-2}\int^{\infty}_{0}dr\, \frac{r^{D-2}}{\sqrt{f(r)}}~,\\
 &=\Omega_{D-2}\int^{\infty}_{0}dr\, \frac{r^{D-2}}{\sqrt{1+\frac{r^2}{l^2}+\frac{\mu}{r^{D-3}}}}~.
 \end{align*}
 Evidently the bulk volume is an IR divergent quantity, and so we impose a bulk IR cutoff, $\Lambda$. The regulated volume complexity turns out to be
 \begin{equation}
 \mathcal{C}_V(T)=\frac{\Omega_{D-2}}{G_N l}\int^{\Lambda}_{0}dr\, \frac{r^{D-2}}{\sqrt{1+\frac{r^2}{l^2}+\frac{\mu}{r^{D-3}}}}~. \label{CV nmSAdS}
 \end{equation}
The integral can only be solved perturbatively in the mass parameter. However, for our purpose, the numerical evaluation of the integral is better suited to unravel the relevant characteristics.  
We, therefore, list the various values of complexity using complexity volume duality along with appropriate IR cutoffs after fixing the AdS radius to be unity.
 \begin{table}[h]
 	\centering
 	\begin{tabular}{|l|c|c|r|}
 		\hline
 		{$\mu$} & IR cutoff &  $\frac{G_N}{\omega_{D-2}}C_V^{SAdS}$& $\frac{G_N}{\omega_{D-2}}(C_V^{SAdS}-C_V^{AdS})$\\
 		\hline
 		\hline
 		2$\times 10^4$ &$ 10^6$ &$ 5.\times 10^{11}$&-3.73\\
 		\hline
 		2$\times 10^5$ &$ 10^6$ & $5.\times 10^{11}$&-37.32\\
 		\hline
 		2$\times 10^6$& $10^6$ & $ 5.\times 10^{11}$& -373.23\\
 		\hline
 		2$\times 10^6$& $10^8$ & $5.\times 10^{15}$&-4\\
 		\hline
 		2$\times 10^7$ & $10^8$ &$5.\times 10^{15}$&-37\\
 		\hline
 		2$\times 10^8$ & $10^8$ & $5.\times 10^{15}$&-373\\
 		\hline
 	\end{tabular}
 \caption{Table for dependence upon mass and cut off for negative mass SAdS using CV}
 \label{CV for negative mass SAdS}
 \end{table}
 The following interesting features are evident from the table \ref{CV for negative mass SAdS} that the volume complexity.
\begin{itemize}
    \item First of all, just like action complexity, the volume complexity displays strong quadratic dependence upon the bulk IR cutoff. As alluded to before, this behavior is expected of any local field theory extensive quantity.
     \item Secondly, the volume complexity display extremely weak dependence on the negative mass parameter $\mu$, the behaviour reminiscent of the action complexity. Recall that this is because the dependence on $\mu$ is heavily suppressed by powers of the large bulk IR cutoff. The fourth column in the table \ref{table:1} displays the values of complexity after background subtraction. Notably, in the last column, it is seen that the difference between the volume complexity of negative mass SAdS and the pure AdS geometry is negative and is orders of magnitude smaller than the pure AdS complexity.   
     \item From the entries of the last column in table (\ref{CV for negative mass SAdS}) one can see that the volume complexity registers a greater difference in the complexity of the negative mass SAdS from the empty AdS that is orders larger than accounted for by the action complexity. This is an artifact of working with $D=4$ as shown in the perturbative analysis performed in Appendix \ref{App.A}. The action complexity leading term has a prefactor $D-4$ while the volume complexity has no such prefactor. As a result the first nontrivial correction in action complexity is of subleading order compared to volume complexity. Disagreements in the subleading pieces of various prescriptions of holographic complexity has been noted widely in the literature for a large class of asymptotically AdS geometric - the underlyig intuition is that these disagreements represent a scheme dependence in their field theory definitions. 
    \item And finally, in marked contrast to the previous case of the action complexity, we are tempted to conjecture the overall negative value after background subtraction as the indication of the sickness of the boundary dual in sync with the Gubser criterion. 
   
\end{itemize}

 Since in the test case of the negative mass SAdS geometry, the CA and CV results could not provide any conclusive complexity criterion for non-allowable naked timelike singularities, in the subsequent sections, we consider two further different concrete examples, to explore further if there can be any possible complexity criterion (akin to the Gubser criterion) which might serve as an alternative diagnostic for the curability of a singular semiclassical geometrical background in a QG theory. 
  
 \section{Timelike Kasner-AdS spacetime} \label{TKAdS}
\begin{figure}[h]
\begin{center} 
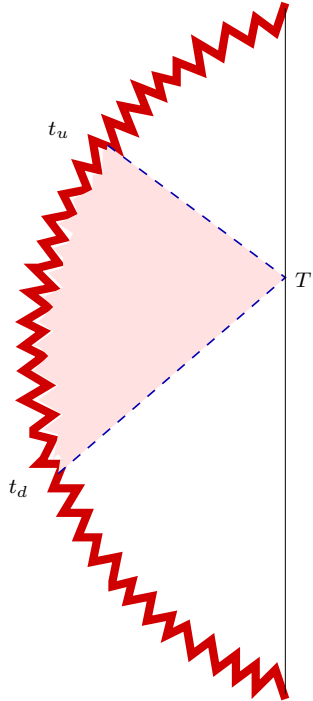
\end{center}
\caption{Penrose diagram for timelike Kasner AdS. The region shaded in pink is the WdW patch corresponding to boundary time $T$.}
	\label{fig2}
\end{figure}

 In this section, we will take a look at an anisotropic solution to the Einstein field equations that appears in effective holographic theories. Effective holographic theories arise in string theory as a result of taking a sequence of limits in which we turn off the stringy physics by setting $\alpha'\to0$ and also the volume of the compactified dimensions is also taken to be zero. This solution is closely related to the more familiar Kasner singularity. However, in this case, the time direction is also allowed to scale anisotropically along with other transverse coordinates. In this work, we are only subjecting the (3+1) dimensional timelike spacetime to treatment. 
The metric for a ``\emph{Wick rotated}'' version of the AdS-Kasner geometry in Poincar\'{e} like conformal coordinates is \cite{Ren:2016xhb}
\begin{equation}
ds^2 = \frac{l^2}{z^2} \left(\frac{dz^2}{f(z)}  - f^{\alpha}(z) \,dt^2 +  f^{\beta}(z)\,dx^2 + f^{\gamma}(z)\,dy^2 \right),\label{Timelike Kasner Metric}
\end{equation}
where $l$ is the AdS radius, and $f(z) = 1 - \frac{z^3}{z_{0}^3}$. We will set $z_{0}=1$ for convenience.
Here exponents $\alpha, \beta, \gamma$ are positive by convention and they satisfy the usual Kasner condition(s)
\begin{equation}\nonumber
\alpha + \beta + \gamma = {\alpha}^2 +{\beta}^2 + {\gamma}^2= 1.
\end{equation}
 This metric has singularities at $z=0$ and $z=1$. While $z=0$ is the spatial infinity (nonsingular), $z=1$ is a \textit{naked} timelike singularity i.e. the metric cannot be continued beyond $0\leq z \leq 1$. The geometry of the singularity at $z=1$ can be more clearly seen by the IR geometry
\begin{equation*}
ds^2 = - r^{2\alpha} dt^2 + dr^2 + r^{2\beta} dx^2 + r^{2\gamma} dy^2~.
\end{equation*}
 The Penrose diagram of this geometry is provided in figure \ref{fig2}.
The limit $\alpha = 1, \beta = \gamma= 0$ reproduces the (isotropic) black brane singularity at $z=1$, so this metric represents the metric for an anisotropic black brane. By taking near boundary limit it is easy to see that geometry (\ref{Timelike Kasner Metric}) connects the AdS at the UV to the timelike Kasner singularity at the IR via RG flow. Despite the fact that the cloaked singularities in anisotropic spacetimes do not adhere to the Gubser criterion because they do not exist in pure gravity \cite{Iizuka:2012wt,Glorioso:2015vrc} or in other words, cannot be obtained as the extremal limit of finite temperature black holes. However, there have been efforts to justify the existence of the anisotropic timelike Kasner singularity can be justified on other grounds, say for example using bulk causality constraints \cite{Kleban:2001nh,Bak:2004yf,Gao:2000ga}. Therefore, if Gubser criterion is to be trusted, we anticipate that the holographic complexity of the timelike Kasner should display some pathological trait. Armed with this belief, we will first tackle the question of holographic complexity using the complexity action proposal and in the subsequent section, try to provide additional justification using the complexity volume prescription. 
\subsection{Action complexity for timelike singularity in Kasner spacetime}
In this section, our aim is to compute the action complexity of the Kasner spacetime featuring naked timelike singularity. The first step towards achieving that is the determination of the WdW patch.  Just like in the previous section, we follow the null rays emanating from the boundary point located at the fixed time $T$. The future (past) null boundaries of the $WdW$ patch are given by the null surfaces,

\begin{equation}
t_{u/d}(z) = T \pm \int_{0}^{z} \frac{dz'}{f^{\frac{1+\alpha}{2}} (z')}. \label{WdW null boundaries}
\end{equation}
where $t_{u/d}$ are respectively the upper/lower limits of the temporal integral.
Following are the various contributions to the on-shell action for the timelike Kasner geometry computed over the WdW patch
\begin{itemize}
\item \textbf{The Einstein-Hilbert bulk contribution}\\
Plugging the integrand $R-2\lambda=-\frac{6}{l^2}$,  the bulk volume contribution to the on-shell action evaluates to
\begin{equation}
   \begin{split}\label{EH-Kasner}
I_{EH}& =\frac{1}{16 \pi G_N}\int_{WdW} \sqrt{g}\left(R-2\lambda\right)~,\\
& =  -\frac{3\,l^2\,V_{xy}}{4\pi G_N} \int_{0}^{1}\frac{dz}{z^4} \int_{0}^{z} \frac{dz'}{f^{\frac{1+\alpha}{2}}(z')}~.
 \end{split}
\end{equation}
Evidently, the Einstein-Hilbert term is independent of the boundary time $T$ and does not make a contribution to the time rate of change of complexity. It diverges as $z\rightarrow0$ and we regulate this divergence by regulator prescription whereby we deform the null boundaries of the WdW patch to timelike, refer to the GHY term calculation in Sec. \ref{the regulator} for the details. The regulated EH term contribution to action complexity is, \begin{align}\label{Kasner EH}
I_{EH}=&-\frac{l^2\,V_{xy}}{8 \pi G_N  \delta ^2} -\frac{3 \,l^2 \,V_{xy}\,\Gamma \left(\frac{4}{3}\right) \sec \left(\frac{\pi  \alpha }{2}\right)}{8\, G_N \,z_0^2\, \Gamma \left(\frac{5}{6}-\frac{\alpha }{2}\right) \Gamma \left(\frac{\alpha -1}{2}\right)}+O(\delta)~.
\end{align}
Here we have restored $z_0$ in the final expression for dimensional consistency. 

\item \textbf{GHY term contribution from the singularity}

The naked singularity at $z=1$ supplies the timelike boundary to the WdW patch and its contribution is given by a suitable GHY term. The pull-back metric of the surface $z=constant,$ 
\begin{equation*}
ds^2 = \frac{L^2}{z^2} \left( - f^{\alpha}(z) \,dt^2 +  f^{\beta}(z)\,dx^2 + f^{\gamma}(z)\,dy^2 \right),
\end{equation*}
from which the volume element can be computed to be $$ d^3x\,\sqrt{h}= l^3\,dt\,dx\,dy\,\frac{ f^{\frac{1}{2}}(z)}{z^3}~.$$ 
The spacelike unit outward normal to this surface has only one non-zero component, $n^z= \frac{z f^{\frac{1}{2}}(z)}{L}~.$ 



Therefore, the trace of the extrinsic curvature turns out to be
\begin{align*}
K &=  \partial_z n^z + \Gamma^X_{Xz} n^z ~,\\&= \frac{z f^{-\frac{1}{2}}(z)f'(z)}{2L}-\frac{3 f^{\frac{1}{2}}(z)}{L}~.
\end{align*}
Thus, the GHY contribution to the action complexity turns out to be
\begin{equation}
\begin{split}\label{Kasner GHY1}
I_{GHY1}&=\frac{1}{8\pi G_N}\,\sqrt{h} K \Big|_{z\rightarrow1} \int dx\,dy\,\int_{t_d}^{t_u}dt ~,\\
& = -\frac{3l^2\,V_{xy}}{8 \pi G_N} \int_0^1 \frac{dz}{f^{\frac{1+\alpha}{2}}(z)}~,\\
& =   -\frac{3 \,l^2\, V_{xy} \,\Gamma \left(\frac{4}{3}\right) \Gamma \left(\frac{1}{2}-\frac{\alpha }{2}\right)}{8 \,\pi \, G_N \, z_0^2\, \Gamma \left(\frac{5}{6}-\frac{\alpha }{2}\right)}+O(\delta)~.
    \end{split}
\end{equation}
Here we have restored $z_0$ in the final result. Thus the contribution from the naked timelike singularity of the Kasner-AdS geometry is finite and negative definite for all $\alpha$ in the allowed range $0<\alpha<1$.

\item \textbf{GHY term from the null boundaries of the WdW patch} \label{the regulator}

The future and past null boundaries of the WdW patch in this case are given by
\begin{align}
\left( t-T \right)^2 - g(z)^2=0\,\,\, \quad \quad\text{where}\quad
g(z)\equiv \int_{0}^{z} \frac{dz'}{f^{\frac{1+\alpha}{2}} (z')}~, \label{WdW bdry surface}
\end{align}
For the future (past) null component $\mathcal{N}_+$ ($\mathcal{N}_-$), choosing the null generator to be $\lambda_+=\frac{-l}{z}$ $\left(\lambda_-=\frac{l}{z}\right)$ gives the components of the null normal to be 
\begin{align}
     k_{\pm}&=\left(\frac{z^2}{l(1-\frac{z^3}{a^3})^{(1+\alpha)/2}}, \mp \frac{z^2}{l},0,0 \right)
\end{align}
Writing out the geodesic equations for the null components measures the deviation from affine parametrization in terms of the parameter measuring departure from affinity,
\begin{equation}
\kappa_{\pm}=\mp\frac{3 (1-\alpha )  z^4}{2l \left(z_0^3-z^3\right)}.    
\end{equation}
The GHY terms from the nill boundary of the WdW patch is then,
\begin{align}
    I_{GHY}^{null} &=-\frac{1}{8 \pi G_N}\int_{\mathcal{N}_+} dS d \lambda_+ \,\kappa_+ + \frac{1}{8 \pi G_N}\int_{\mathcal{N}_-} dS d \lambda_- \,\kappa_- \nonumber\\
    &=\frac{3V_{xy}l^2}{8 \pi G_N z_0^2}\frac{(1-\alpha)\Gamma \left(\frac{4}{3}\right) \Gamma \left(\frac{1-\alpha}{2}\right)}{\Gamma \left(\frac{1}{6} (5-3 \alpha )\right)} \label{Kasner GHY}
\end{align}

\item \textbf{LMPS counterterms} 
The pullback metric on the constant $\lambda_+$ sections on the future null boundary of WdW patch $\mathcal{N}_+$ is 
\begin{equation*}
 dh_+^2=\frac{l^2}{z^2} \left(  f^{\beta}(z)\,dx^2 + f^{\gamma}(z)\,dy^2 \right)
\end{equation*}
Then,
\begin{align}   
    \Theta_+&\equiv\frac{1}{\sqrt{h_+}}\frac{\partial \sqrt{h_+}}{\partial \lambda_+}=-\frac{z \left((\alpha -1) z f'(z)+4 f(z)\right)}{2 l f(z)}\nonumber\\
    &=\frac{z \left((3 \alpha +1) z^3-4 z_0^3\right)}{2 l \left(z_0^3-z^3\right)}
\end{align}
The null component $\mathcal{N}_+$ lies to the future of the bulk of WdW therefore, sign$(\mathcal{N}_+)=+1$, and with this choice,
\begin{align}
I_{LMPS+}&=-\frac{1}{8 \pi G_N}\int_{\mathcal{N}_+}dS d \lambda_+ \Theta_+\log|\Theta_+|\nonumber \\
&=\frac{l^3 \,V_{xy}}{8 \pi G_N}\int_{\delta}^{z_0}dz\,\frac{1}{z^4}\left(1-\frac{z^3}{z_0^3}\right)^{\frac{1-\alpha}{2}} \frac{z \left(4 z_0^3-(3 \alpha +1) z^3\right)}{2 l \left(z_0^3-z^3\right)}\ln\left| \frac{z \left((3 \alpha +1) z^3-4 z_0^3\right)}{2 l \left(z_0^3-z^3\right)}\right| 
\end{align}
{For the past null boundary of the WdW patch, $\mathcal{N}_-$, sign($\mathcal{N}_-)=-1$ because $\mathcal{N}_-$ lies to  the past of the bulk of WdW, and 
\begin{align*}
    \sqrt{h_-}&=\sqrt{h_+}, 
    \lambda_-=-\lambda_+,
    \Theta_-=-\Theta_+
\end{align*}
Then,
\begin{align}   
I_{LMPS-}&= \frac{1}{8 \pi G_N}\int_{\mathcal{N}_-}dS d \lambda_- \Theta_-\log|\Theta_- \nonumber\\
& = I_{LMPS+} \nonumber
\end{align}
So the full LMPS term contribution,
\begin{align}
    I_{LMPS} &= I_{LMPS+}+I_{LMPS-} \nonumber\\
     & =\frac{l^2 V_{xy}}{4 \pi G_N \delta ^2}\left(-\ln\left(\frac{l }{2 \delta}\right)+\frac{1}{2}\right)+ \frac{l^2 V_{xy}}{ 8 \pi G_N z_0^2} \left(\frac{3+\alpha}{2}\right) \frac{\Gamma \left(\frac{1}{3}\right) \Gamma \left(\frac{1-\alpha}{2}\right)}{\Gamma \left(\frac{5}{6} - \frac{\alpha}{2} \right)} \label{Kasner LMPS}
\end{align}
\item \textbf{Joint terms from singularity and WdW null boundary}}
Here we compute the contribution from edges resulting from the intersection/joint of the WdW boundary and the singularity. To this end we need the inner product of the normals:
\begin{align}
    k_+\cdot n&=\frac{l^2}{z^2f(z)}\frac{zf(z)^{1/2}}{l}\frac{z^2}{l}=\frac{z}{f(z)^{1/2}}\\
    k_-\cdot n &=-k_+\cdot n= \frac{-z}{f(z)^{1/2}}\\
    k_+\cdot k_-&=\frac{-2z^2}{\left(1-\frac{z^3}{z_0^3}\right)}
\end{align}
For intersection of the timelike singularity with the WdW null boundary (future or past)
\begin{align}
    I_{j_{\pm}}&=\lim_{z\to z_0} \frac{1}{8 \pi G_N}\int  dx\,dy \,\frac{l^2}{z^2}\left(1-\frac{z^3}{z_0^3}\right)^{\frac{1-\alpha}{2}}\ln\left|\frac{k_{\pm}\cdot n}{l}\right| \nonumber\\
    &=\lim_{z\to z_0} \frac{V_{xy}}{8 \pi G_N}\frac{l^2}{z^2}\left(1-\frac{z^3}{z_0^3}\right)^{\frac{1-\alpha}{2}}\log\left|\frac{z}{l\left(1-\frac{z^3}{z_0^3}\right)^{1/2}}\right| \nonumber \\& = 0.
\end{align}
Thus both the null-timelike joints offer vanishing contribution.\\
Next consider the contribution from the intersection of the future and past boundaries of the WdW patch, the so called \textbf{null-null Joint: }
\begin{align}
   I_{nn}&=-\lim_{z\to \delta}\, \frac{1}{8 \pi G_N}\int dx dy \,\frac{l^2}{z^2}\left(1-\frac{z^3}{z_0^3}\right)^{\frac{1-\alpha}{2}}\log\left|\frac{k_{\pm}\cdot k_-}{2l}\right| \nonumber \\
    &=  \frac{V_{xy} l^2}{4 \pi G_N \delta^2}\ln\left(\frac{l}{\delta}\right). \label{Kasner nn}
\end{align}
\end{itemize}

We now combine (\ref{Kasner EH}), (\ref{Kasner GHY1}) \eqref{Kasner GHY}, \eqref{Kasner LMPS}, and \eqref{Kasner nn} and plug them back in (\ref{CA}) to obtain the following closed form expression for the action complexity of timelike Kasner-AdS spacetime (in natural units)
\begin{align}\label{Kasner CA}
    \mathcal{C}_A &= \frac{l^2  \ln2}{4 \pi^2   G_N}\frac{V_{xy}}{\delta^2}+\frac{l^2 }{8 \pi^2 G_N}\frac{V_{xy}}{z_0^2}\frac{\left(3 \pi  (2-\alpha ) \Gamma \left(\frac{4}{3}\right) \sec \left(\frac{\pi  \alpha }{2}\right)\right)}{ \Gamma \left(\frac{5}{6}-\frac{\alpha }{2}\right) \Gamma \left(\frac{\alpha +1}{2}\right)}~.
\end{align}
The leading divergent piece corresponds to the pure AdS$_4$ contribution to the action complexity in the Poincar\'e patch  \cite{Reynolds:2016rvl} and the finite term comes from the deformation to Kasner-AdS. This result is not surprising because the bulk IR (boundary UV) divergence arises from the bulk IR region where the geometry is locally pure AdS$_4$ and the Kasner singularity makes an impact only in deep interior of the bulk (boundary IR). We note that the finite term is a monotonically increasing function of the Kasner exponent $\alpha$ and and is positive definite, i.e. \emph{timelike Kasner has higher action complexity than empty Poincar\'e AdS$_4$}. This trait is similar to that of negative mass Schwarzschild-AdS, which had higher complexity than global AdS and thus there are no pathological features diagnosable in the action-complexity expression. This analysis hints that the action-complexity might not be a suitable probe for allowable timelike singularities (independent of Gubser criterion). However, there is an alternative prescription, namely the volume complexity and it would be interesting to see whether the volume complexity prescription also leads to the same conclusion. With this in mind, we proceed in the upcoming subsection to compute the volume complexity of the field theory dual to the timelike Kasner-AdS spacetime. 

\subsection{Volume complexity of timelike Kasner-AdS}
Consider the general codimension-one spatial slice anchored at the boundary time, $T$ is obtained by treating $t=t(z)$. Then, the induced metric on the hypersurface becomes
\begin{equation}\nonumber
ds^2=\frac{L^2}{z^2}\left(\left(-f^{\alpha}t'^2+\frac{1}{f}\right)dz^2+f^{\beta}dx^2+f^{\gamma}dy^2\right)~.
\end{equation}
Therefore, we will extremize the following volume
\begin{equation}
V=V_{xy}\int_0^1dz\,\frac{l^3
}{z^3}\sqrt{\left(-f^{\alpha}t'^2+\frac{1}{f}\right)f^{1-\alpha}}~.\label{TtKV}
\end{equation}
The Euler-Lagrange equation imposing the condition of maximality upon the spatial volume is
\begin{equation}\nonumber
2 z \left(z^3-1\right) t''(z)+3 \left(\alpha  z^3+2\right) t'(z)-3 \left(z^6-3 z^3+2\right) \left(1-z^3\right)^{\alpha } t'(z)^3=0~.
\end{equation}
Assuming perturbative expansion about the boundary $z=0, t=T$, and solving term by term leads us to the following constant time slice as the solution
\begin{equation}\nonumber
t(z)= T~.
\end{equation}
This solution is obviously the maximum since any nonzero derivative $t'$ in the volume functional \eqref{TtKV} would lower the volume of the hypersurface. So the maximum volume is attained when $t'=0$, i.e. $t=T$. According to the volume-complexity prescription, the volume complexity of the boundary state at time $t=T$ is
\begin{equation}
\begin{split}\label{Kasner CV}
\mathcal{C}_V&=
\frac{l^2V_{xy}}{G_N}\int_{\delta}^1dz\,\frac{1
}{z^3 f(z)^{\frac{\alpha}{2}}}~,
\\&=\frac{l^2}{2\, G_N}\frac{ V_{xy}}{\delta ^2}-\frac{l^2 }{2\,G_N}\frac{V_{xy}}{z_0^2}\frac{\Gamma \left(\frac{1}{3}\right) \Gamma \left(\frac{2-\alpha }{2}\right) }{ \Gamma \left(\frac{2-3\alpha }{6}\right) }+O(\delta)~.
\end{split}
\end{equation}
As expected, the volume complexity has a leading quadratic UV divergence that is characteristic of a local field theory with $2$ spatial dimensions dual to the pure AdS$_{3+1}$ bulk. Here too this leading divergent piece is independent of the Kasner exponent(s).  Volume complexity also has a finite piece, which is also a monotonically increasing function of the Kasner exponent, $\alpha$, similar to the  action complexity case. However, the finite pieces of the volume and action complexities are distinct (even after taking into account the mismatch of the leading quadratic divergent pieces). This, however, is not unexpected - the mismatch between the subleading terms in the volume and action complexity has been well documented in the literature \cite{Bolognesi:2018ion, Chakraborty:2020fpt,Katoch:2022hdf}. Compared to the action-complexity result, where the finite piece was positive definite, here we obtain a finite piece that shows a change on sign from the negative to the positive values across $\alpha=2/3$. Therefore, the volume complexity results are in sync with what the Gubser criterion only for a somewhat restricted range of the Kasner coefficients, $\alpha<2/3$.

So far, from the two cases examined, we have not been able to find any suitable complexity criterion (be it volume or action) which can be used to diagnose allowable timelike naked singularities akin to the Gubser criterion. We consider one more case of an asymptotically AdS geometry with a naked timelike singularity in the next section. The geometry, obtained as a solution to Einstein-Maxwell-Dilation system in asymptotically AdS, exhibits an interesting phase transition between the allowed phase and the disallowed phase as per the Gubser criterion. Our aim is twofold, firstly we hope to see the complexity too undergoing an analogous phase transition, from the sick geometrical phase to the allowed phase. And in due process, we also hope to settle the issue of whether complexity (and in particular which prescription) can be a suitable tool to probe the nature of geometries with naked timelike singularities - can it inform us which naked timelike singularities are allowed in the underlying theory or not. 
 
\section{The Einstein-dilaton system: Singularities \& Gubser criterion} \label{ED}
We will begin this section by reviewing the Gubser criterion for the planar black holes arising in the Einstein-dilaton system. It is known that spacetime singularities in the IR may or may not be acceptable in terms of the AdS/CFT correspondence. Consider an Einstein-dilation system described by the action
\begin{equation}
	I=\frac{1}{16\pi G_N} \int d^{4}x\sqrt{-g}\left[
	R-\frac{1}{2}(\partial_\mu\phi)^2-V(\phi)\right]~,\label{EDaction}
\end{equation}
with a given potential $V(\phi)$. The IR geometry may contain a naked singularity. Gubser~\cite{Gubser:2000nd} proposed a criterion to distinguish acceptable and unacceptable singularities. The Gubser criterion has the following two statements:
\begin{itemize}
\item[(A)] The potential of the scalar field is bounded from the above in the solution.
\item[(B)] The spacetime can be obtained as a limit of a regular black hole. This is a weaker form of the cosmic censorship principle.
\end{itemize}
The two statements are not exactly equivalent, but it was argued in~\cite{Gubser:2000nd} that statement (A) often implies statement (B). If statement (B) is satisfied, the singularity is good, i.e., acceptable.

The Einstein-scalar system with the following potential
\begin{equation}
V(\phi)=-\frac{2}{(1+\alpha^2)^2l^2}\Bigl[\alpha^2(3\alpha^2-1)e^{-\phi/\alpha}
+8\alpha^2e^{(\alpha-1/\alpha)\phi/2}+(3-\alpha^2)e^{\alpha\phi}\Bigr],\label{eq:potential}
\end{equation}
enjoys an analytic solution, see \eqref{ED_metric}. A nice feature of this potential is that the special values of the parameter $\alpha=\frac{1}{\sqrt{3}}, 1, \sqrt{3}$, correspond to special cases in STU supergravity. The $\phi\to 0$ behavior is 
\begin{equation*}
V(\phi)=-6/l^2-(1/l^2)\phi^2+\ldots,
\end{equation*}
where the first term is a cosmological constant. We assume $\alpha>0$ throughout.

This potential together with an exact solution of an Einstein-Maxwell-Scalar system was found in~\cite{Gao:2004tu}. A nontrivial neutral limit of this solution gives an Einstein-dilaton system with naked spacetime singularities in the IR~\cite{Ren:2019lgw}. The solution of the metric and the scalar field is given by
\begin{gather}
ds^2=f(r)(-dt^2+d\vec{x}^2)+f^{-1}(r)dr^2, \label{ED_metric}\\
f(r)=\frac{r^2}{l^2}\left(1+\frac{Q}{r}\right)^\frac{2}{1+\alpha^2},\qquad \phi=-\frac{2\alpha}{1+\alpha^2}\ln\left(1+\frac{Q}{r}\right).\nonumber
\end{gather}
The Kretschmann scalar for this metric is
\begin{equation}\nonumber
R_{\mu\nu\rho\sigma}R^{\mu\nu\rho\sigma}=\frac{12 ((r+(r+Q) \alpha ^2)^4+(r-Q \alpha +(r+Q) \alpha ^2)^2 (r+Q \alpha +(r+Q) \alpha ^2)^2)}{l^4 \,(1+\alpha ^2)^4\, (r+Q)^{\frac{4 \alpha ^2}{1+\alpha ^2}} \,r^{\frac{4}{1+\alpha ^2}}}~.
\end{equation}
There is a curvature singularity at $r=0$ and $r=-Q$. The region we are interested in is between the AdS boundary (UV) and the spacetime singularity (IR). The AdS boundary is at $r\to\infty$. When $Q>0$, there is a spacetime singularity at $r=0$ in the IR. When $Q<0$, there is a spacetime singularity at $r=-Q$ in the IR.

Consider the $Q>0$ case first. There is a spacetime singularity at $r=0$. We will use the Gubser criterion to examine whether this singularity is acceptable, and show that the two statements give the same range of the parameter $\alpha$, except for the marginal case, $\alpha=1/\sqrt{3}$.

For statement (A), we plug the solution of $\phi$ in the potential, and obtain
\begin{equation}\nonumber
V=-\frac{2 \alpha ^2 (3 \alpha ^2-1) Q^2+12 \alpha ^2 (1+\alpha ^2) Q r+6 (1+\alpha ^2)^2 r^2}{l^2 (1+\alpha ^2)^2  (r+Q)^{\frac{2 \alpha ^2}{1+\alpha ^2}}r^{\frac{2}{1+\alpha ^2}}}.
\end{equation}
The divergence of the potential is at the IR, $r\to 0$. The $r\to 0$ behavior is
\begin{equation}\nonumber
\lim_{r\to 0} V(\phi)\to
\begin{cases}
+\infty,\qquad \alpha<1/\sqrt{3}\,,\\\nonumber
-\infty,\qquad \alpha\geq 1/\sqrt{3}\,.\nonumber
\end{cases}
\qquad (Q>0)
\end{equation}
According to statement (A), the singularity at $r=0$ is acceptable if $\alpha\geq 1/\sqrt{3}$.

For statement (B), we need to find a near-extremal geometry, which is at finite temperature and the horizon encloses the singularity. If such a geometry exists, the singularity is acceptable or resolvable. The analytic solution of a finite temperature black hole in the above Einstein-dilaton system is not available. However, it is sufficient to examine the IR geometry. To obtain the IR geometry, we take the $r\to 0$ limit and keep the leading term. The potential is
\begin{equation}\nonumber
V= -V_0\,e^{-\phi/\alpha}~,
\label{eq:V-IR}
\end{equation}
where $V_0=\frac{2\alpha^2(3\alpha^2-1)}{(1+\alpha^2)^2L^2}$.
For this potential, the analytic solution as the IR geometry is
\begin{align}\nonumber
ds^2 &=-f_0r^\frac{2\alpha^2}{1+\alpha^2}dt^2+\frac{dr^2}{g_0r^\frac{2\alpha^2}{1+\alpha^2}}+f_0r^\frac{2\alpha^2}{1+\alpha^2}d\vec{x}^2,\\
\phi &=\ln\left(e^{\phi_0}r^\frac{2\alpha}{1+\alpha^2}\right)~,\nonumber
\end{align}
where $f_0=g_0=Q^{\frac{2}{1+\alpha^2}}/L^2$ and $e^{\phi_0}= Q^{\frac{-2\alpha}{1+\alpha^2}}$.
This is a hyperscaling-violating geometry with the Lifshitz scaling $z=1$. When $\alpha>1$, the singularity at $r=0$ is null. When $\alpha<1$, the singularity at $r=0$ is timelike. To see this, we can write the metric near $r=0$ as
\begin{equation}
ds^2\sim -\rho^2dt^2+\rho^\frac{2(1-\alpha^2)}{\alpha^2}d\rho^2+\rho^2d\vec{x}^2\,,\nonumber
\end{equation}
where $\rho\sim r^\frac{\alpha^2}{1+\alpha^2}$. A normal covector of constant $\rho$ is $n_\mu=\nabla_\mu\rho=(0,1,0,0)$. Then we have $n^\mu n_\mu=g^{\rho\rho}\sim\rho^\frac{2(\alpha^2-1)}{\alpha^2}$. When $\alpha>1$, we have $n^\mu n_\mu=0$ at the singularity. When $\alpha<1$, we have a (naked) timelike singularity, which is the case of interest for us.\\

By adding a blackening factor, we obtain a near-extremal geometry given by
\begin{equation}\label{IR geometry}
ds^2=-f_0r^\frac{2\alpha^2}{1+\alpha^2}\left[1-\left(\frac{r_h}{r}\right)^\beta\right]dt^2+\frac{dr^2}{g_0r^\frac{2\alpha^2}{1+\alpha^2}\left[1-\left(\dfrac{r_h}{r}\right)^\beta\right]}+f_0r^\frac{2\alpha^2}{1+\alpha^2}d\vec{x}^2~,
\end{equation}
where $r_h$ is the horizon radius and
\begin{equation}\nonumber
\beta=\frac{3\alpha^2-1}{1+\alpha^2}~.
\end{equation}
This is an exact solution to the Einstein-dilaton system with the potential~\eqref{eq:potential}. The range of $r$ is $r_h<r<\infty$, so $r_h/r<1$.  If this solution makes sense for describing a black hole, we need $\beta>0$, which gives
\begin{equation}\nonumber
\alpha>1/\sqrt{3}~.
\end{equation}
Namely, $\alpha\leq 1/\sqrt{3}$ violates the Gubser criterion. If we ignore the marginal case $\alpha=1/\sqrt{3}$, the statements (A) and (B) are consistent.

In the next and subsequent section, we present the details of the action complexity and volume complexity for the $Q>0$ case for the hyperscaling violating black holes with timelike singularity in the IR. 

\subsection{Action complexity for Einstein-Scalar system}
In this section, we tackle the problem of governing the holographic complexity of the Einstein-dilaton system using action complexity conjecture. As alluded to in the above section, as $\alpha$ is dialed from below $\alpha=1/\sqrt{3}$ to the higher values, the naked timelike singularity turns from an unacceptable type to an acceptable type according to the Gubser criterion. If complexity is to be a reliable probe of timelike singularities, then both complexity prescriptions are expected to successfully capture this feature by showing some abrupt change (phase transition) in their behavior, and from the features of complexity for $\alpha<1/\sqrt{3}$\, one should be able to formulate a complexity criterion for signalling unacceptable naked timelike singularities.

Einstein-dilation system is described by the gravitational action in (\ref{EDaction}).
For the sake of calculational convenience, we treat each term of the potential individually (\ref{eq:potential}) by writing, $V(\phi)=V_1(\phi)+V_2(\phi)+V_3(\phi)$ where 
\begin{equation}\nonumber
 V(\phi)=\frac{2 \alpha ^2 \left(1-3 \alpha ^2\right)}{\left(\alpha ^2+1\right)^2 l^2}\left(1+\frac{Q}{r}\right)^{\frac{2}{\alpha ^2+1}}-\frac{16 \alpha ^2 }{\left(\alpha ^2+1\right)^2 l^2}\left(1+\frac{Q}{r}\right)^{\frac{1-\alpha ^2}{\alpha ^2+1}}-\frac{2 \left(3-\alpha ^2\right)}{\left(\alpha ^2+1\right)^2 l^2}\left(1+\frac{Q}{r}\right)^{-\frac{2 \alpha ^2}{\alpha ^2+1}}~.
\end{equation} 

The analytic solution to the above action is supplied by the geometry in \eqref{ED_metric}, whose  Ricci scalar can be computed to give
\begin{equation}\nonumber
R=-\frac{6\left(2 \left(\alpha ^2 (Q+r)+r\right)^2-\alpha ^2 Q^2\right)}{l^2\left(\alpha ^2+1\right)^2 (Q+r)^2} \left(\frac{Q+r}{r}\right)^{\frac{2}{\alpha ^2+1}} ~.
\end{equation}
As usual, the Wheeler-de Witt patch is obtained by following the evolution of light rays towards past and future, starting from the boundary at a given time and can be found to be bounded by the future (past) light-sheets
\begin{align*}
t_{\pm}(r)&=t_{*}\mp\int_{\infty}^{r}\frac{dr'}{f(r')}~.
\end{align*}
We will first present the evaluation of the bulk action terms and in the subsequent subsections present the details of the higher codimension contributions.
\subsubsection{Complexity contributions from the bulk}
We split the on-shell bulk action into the sum of different contributions in the following manner
\begin{equation}\nonumber
	I_{bulk}=I_{EH}+I_{T(\phi)}-I_{V}~.
\end{equation}
Here the first two terms correspond to the Einstein-Hilbert term and scalar kinetic energy term, and the last three correspond to the potential terms for the scalar. All these contributions diverge as $r\rightarrow \infty$ and we regulate of the contributing integrals with the aid of the IR cutoff, $\Lambda$. Refer to Sec.~\ref{GHY_WdW_ES} for the details of the regularization procedure. We list the bulk contributions in the following:
\begin{itemize}
    \item \textbf{Einstein-Hilbert term:}
   \small{ \begin{align}\label{ED EH}
        I_{EH}&=\frac{1}{16\pi G_{N}}\int dx\,dy\int dr\sqrt{-g}\:R\left(\int_{t_{-}(r)}^{t_{+}(r)}dt\right)\nonumber\\
        &=\frac{-6	V_{xy}}{16\pi G_{N}\left(\alpha ^2+1\right)^2}\int_{0}^{\Lambda}dr\, r^2 \left(\frac{Q+r}{r}\right)^{\frac{4}{\alpha ^2+1}}\frac{ \left(2 \left(\alpha ^2 (Q+r)+r\right)^2-\alpha ^2 Q^2\right)}{ (Q+r)^2}\int_{r}^{\Lambda}\frac{dr'}{r'^2\left(1+\frac{Q}{r'}\right)^{\frac{2}{1+\alpha^2}}}\nonumber\\
        &=\frac{2V_{xy}}{16\pi G_N}\biggl(-\frac{2 \Lambda ^2}{l^2}-\frac{4 \Lambda  Q}{\left(\alpha ^2+1\right) l^2}+\frac{2 \alpha ^2 Q^2 \log \left(\frac{\Lambda }{Q}\right)}{\left(\alpha ^2+1\right)^2l^2}+\frac{6 \alpha ^2 \left(\alpha ^2+1\right) \left(2 \alpha ^2-1\right) \epsilon ^{\frac{3 \alpha ^2-1}{\alpha ^2+1}} Q^{\frac{3-\alpha ^2}{\alpha ^2+1}}}{\left(3 \alpha ^2-1\right) \left(\alpha ^4-1\right) l^2}\nonumber\\
        &\hspace{0.5\textwidth}+(\text{finite})+O(\Lambda^{-1})\biggr).
    \end{align}}
Where we mention the finite term separately for brevity to be 
\begin{multline}
        \frac{2 Q^2}{\left(\alpha ^2+1\right)^2 l^2}\Bigg[-2 \alpha ^4-\alpha ^2+4 \left(1-3 \alpha ^2\right) \alpha ^2 \left(\psi ^{(0)}\left(\frac{4 \alpha ^2}{\alpha ^2+1}\right)+\gamma \right)\\
        +6 \left(2 \alpha ^2-1\right) \alpha ^2 H_{2-\frac{4}{\alpha ^2+1}}+12 \alpha ^4 H_{\frac{2 \alpha ^2}{\alpha ^2+1}}+\left(\alpha ^2-12 \alpha ^4\right) H_{1-\frac{2}{\alpha ^2+1}}-1\Bigg].
\end{multline}
    \item \textbf{Kinetic term for the scalar field:}
    \begin{equation}
        \begin{split}\label{eq: Scalar kinetic term}
        I_{T(\phi)}&=\frac{1}{16\pi G_{N}}\int dx\,dy\int dr\sqrt{-g}\:\left(-\frac{g^{rr}}{2}\left(\partial_{r}\phi\right)^{2}\left(\int_{t_{-}(r)}^{t_{+}(r)}dt\right)\right)~,\\
        &=\frac{-4V_{xy} \alpha ^2 Q^2 }{16\pi G_{N}\left(\alpha ^2+1\right)^2 }
\int_{0}^{\Lambda} dr\,\left(1+\frac{Q}{r}\right)^{\frac{2(1-\alpha^2)}{1+\alpha^2}}\int_{r}^{\Lambda}\frac{dr'}{r'^2\left(1+\frac{Q}{r'}\right)^{\frac{2}{1+\alpha^2}}}~,\\
&=\frac{2V_{xy}}{16\pi G_N}\left(-\frac{2 \alpha ^2 Q^2 \log \left(\frac{\Lambda }{Q}\right)}{\left(\alpha ^2+1\right)^2 l^2}+\frac{2 \alpha ^2 \left(\alpha ^2+1\right) \epsilon ^{\frac{3 \alpha ^2-1}{\alpha ^2+1}} Q^{\frac{3-\alpha ^2}{\alpha ^2+1}}}{\left(3 \alpha ^2-1\right) \left(1-\alpha ^4\right) l^2}+(\text{finite})+O\left(\Lambda^{-1}\right)\right)~,
 \end{split}
    \end{equation}
    the finite term is 
    \begin{equation}
        \begin{split}
            \frac{2 \gamma  \alpha ^2 Q^2}{\left(\alpha ^2+1\right)^2 l^2}&+\frac{4 \alpha ^2 Q^2 \psi ^{(0)}\left(\frac{3\alpha^2-1}{\alpha ^2+1}+1\right)}{\left(\alpha ^2+1\right)^2 l^2}-\frac{2 \alpha ^2 Q^2 \psi ^{(0)}\left(\frac{2 \alpha ^2}{\alpha ^2+1}\right)}{\left(\alpha ^2+1\right)^2 l^2}
        \end{split}
    \end{equation}
    \item \textbf{Scalar potential term in action: }
    \begin{equation}
 \begin{split}
       I_{V(\phi)}&=\frac{1}{ 16\pi G_{N}}\int dx\,dy\int dr\sqrt{-g}\:V(\phi)\:\left(\int_{t_{-}(r)}^{t_{+}(r)}dt\right)~,\\
        &=\frac{2V_{xy}}{ 16\pi G_{N}}\int_{0}^{\Lambda}dr\:r^2\left(1+\frac{Q}{r}\right)^\frac{2}{1+\alpha^2}\:V(\phi)\,\int_{r}^{\Lambda}\frac{dr'}{r'^2\left(1+\frac{Q}{r'}\right)^{\frac{2}{1+\alpha^2}}}~,\\
        &=\frac{2V_{xy}}{16\pi G_N}\left( -\frac{\Lambda ^2}{l^2}-\frac{2 \Lambda  Q}{\left(\alpha ^2+1\right) l^2}-\frac{ 2 \alpha ^2 Q^{\frac{3-\alpha^2}{\alpha ^2+1}} \epsilon ^{\frac{3 \alpha ^2-1}{\alpha ^2+1}}}{\left(\alpha ^2-1\right) l^2} +(\text{finite})+O(\Lambda^{-1})\right)~.\label{eq: Scalar potential term}
 \end{split}
 \end{equation}
 Where the finite term is 
 \begin{multline}
         \frac{Q^2}{\left(\alpha ^2+1\right)^2 l^2} \Bigg[-2 \alpha ^4-\alpha ^2+4 \left(1-3 \alpha ^2\right) \alpha ^2 \left(\psi ^{(0)}\left(\frac{4 \alpha ^2}{\alpha ^2+1}\right)+\gamma \right)\\
         +4 \left(3 \alpha ^2-1\right) \alpha ^2 H_{2-\frac{4}{\alpha ^2+1}}+12 \alpha ^4 H_{\frac{2 \alpha ^2}{\alpha ^2+1}}-12 \alpha ^4 H_{1-\frac{2}{\alpha ^2+1}}-1\Bigg].
 \end{multline}
The integrals are divergent at the lower limit of integration, $\varepsilon$ is the bulk UV cutoff (equivalently the IR cutoff in the boundary dual) at the lower limit near the singularity.
\end{itemize}
\subsubsection{Complexity contribution from the singularity}

The naked timelike singularity is located at $r=0$. The induced metric on fixed
$r$ surface is
\[
ds^{2}=-f(r)\:dt^{2}+f(r)\left(dx^{2}+dy^{2}\right),
\]
while the unit outward normal to the WdW volume is $n=-f^{1/2}(r)\;\partial_{r}$.
These lead to the induced metric determinant
\[
\sqrt{-h}=f^{3/2}(r),
\]
and the trace of the extrinsic curvature
\[
K\equiv\nabla_{L}n^{L}=\partial_{r}n^{r}+\Gamma_{Lr}^{L}n^{r}=-\frac{3}{2}f'(r)f^{-1/2}(r).
\]
The GHY term contribution from the singularity is
\[
16\pi G_{N}\:I_{GHY}^{0}=\lim_{r\rightarrow0}2\int d^{3}x\:\sqrt{-h}\:K=-3V_{xy}\lim_{r\rightarrow0}\left(t_{u}(r)-t_{d}(r)\right)\left(f'(r)f(r)\right)
\]
Using the limiting forms as $r\rightarrow0$,
\[
f(r)\sim Q^{\frac{2}{1+\alpha^{2}}}r^{\frac{2\alpha^{2}}{1+\alpha^{2}}},\qquad t_{u}(r)-t_{d}(r)=\frac{\left(\alpha^{2}+1\right)\left(\frac{Q}{r}\right)^{\frac{\alpha^{2}-1}{\alpha^{2}+1}}}{\left(\alpha^{2}-1\right)Q}
\]
we get,
\begin{equation}
16\pi G_{N}\:I_{GHY}^{0}=-3V_{xy}\frac{2\alpha^{2}}{\alpha^{2}-1}Q^{\frac{2}{1+\alpha^{2}}}\left(\lim_{r\rightarrow0}r^{\frac{2\alpha^{2}}{1+\alpha^{2}}}\right)=0.\label{eq: GHY term from singularity}
\end{equation}
Thus the contribution to action complexity from the naked timelike singularity vanishes!

\subsubsection{Contribution from the null boundary of the WdW patch} \label{GHY_WdW_ES}

The null boundaries of the WdW patch is given by
\begin{align}
    (t-T)^2-\left(\int_r^{\infty}\frac{dr'}{f(r')}\right)^2=0~,
\end{align}
For the future (past) boundary of the WdW patch, $\mathcal{N_+}$ ($\mathcal{N_-}$) we choose the parameterization ,
\begin{align}
\lambda_\pm&=\mp\frac{r}{l}
\end{align}
The corresponding null generators are,
\begin{equation}
k_{\pm}= \left(\frac{l}{f(r)},\mp l,0,0\right),
\end{equation}
for which we obtain
\begin{equation}
\kappa_{\pm} = 0.
\end{equation}
i.e. this choice of null parametrization is affine. Consequently there will be no GHY type contribution from the WdW null boundary.

There is however, the LMPS term contribution. For this we need the metric on the sections of constant $\lambda_\pm$ on $\mathcal{N}_\pm$, namely,
\begin{equation*}
ds^2 = f(r)(dx^2+dy^2).    
\end{equation*}
Then, $\sqrt{h}=f(r)$ and the expansion,
\begin{align}
    \Theta_{\pm}&\equiv \frac{1}{\sqrt{h}}\frac{\partial \sqrt{\gamma}}{\partial \lambda_{\pm}} \nonumber \\
    &=\mp \frac{l}{f(r)}\frac{\partial f(r)}{\partial r}\nonumber\\&=\mp \frac{2 l \left(\alpha ^2 (Q+r)+r\right)}{\left(\alpha ^2+1\right) r (Q+r)}
\end{align}
The LMPS term from $\mathcal{N}_+$
\begin{align}
   I_{LMPS+}&=-\frac{1}{8\pi G_N} \int_{\mathcal{N}_+}dS d\lambda_+\,\Theta_+\log|\Theta_+| \nonumber\\
     &=\frac{4V_{xy}}{16\pi G_N} \int^{\infty}_{0}d\left(\frac{r}{l}\right)\,\left(\frac{Q+r}{r}\right)^{\frac{1-\alpha ^2}{1+\alpha ^2}} \frac{\left(\alpha ^2 (Q+r)+r\right)}{\left(\alpha ^2+1\right) l}\log\left|\frac{2 l \left(\alpha ^2 (Q+r)+r\right)}{\left(\alpha ^2+1\right) r (Q+r)}\right|
\end{align}
while,
\begin{equation*}
    I_{LMPS-}= I_{LMPS+}
\end{equation*}
Therefore, the total LMPS contribution is 
\begin{align}
    I_{LMPS} &= 2I_{LMPS+} \nonumber\\
    &=\frac{4V_{xy}}{8\pi G_N l} \int^{\Lambda}_{0}dr\,\left(\frac{Q+r}{r}\right)^{\frac{1-\alpha ^2}{1+\alpha ^2}} \frac{\left(\alpha ^2 (Q+r)+r\right)}{\left(\alpha ^2+1\right) l}\ln\left|\frac{2 l \left(\alpha ^2 (Q+r)+r\right)}{\left(\alpha ^2+1\right) r (Q+r)}\right| \nonumber \\
    & =  \frac{V_{xy}}{8 \pi G_N} \left[ \frac{\Lambda^2}{l^2}  - \frac{2\Lambda^2}{l^2} \ln\left(\frac{\Lambda}{2l}\right) -\frac{4Q}{l(1+\alpha^2)} \frac{\Lambda}{l} \ln\left(\frac{\Lambda}{2l}\right) -\frac{2\left(1-2 \alpha ^2\right) Q^2 }{\left(1+\alpha ^2\right)^2 l^2}\ln \left(\frac{\Lambda}{2l}\right) + \text{finite}\right] \label{ED LMPS}
\end{align}
where the finite piece is equal to,
\begin{equation}
\frac{Q^2 \left(-2 \left(\alpha ^2-1\right) \alpha ^2 H_{-\frac{2}{\alpha ^2+1}}-2 \left(\alpha ^4+\alpha ^2\right) \, _2F_1\left(1,1-\frac{2}{\alpha ^2+1};\frac{2 \alpha ^2}{\alpha ^2+1};-\frac{1}{\alpha ^2}\right)+\left(\alpha ^2-1\right) \left(\alpha ^2+2 \alpha ^2 \ln\left(\frac{2l}{Q}\right)-2\right)\right)}{\left(\alpha ^2-1\right) \left(\alpha ^2+1\right)^2 l^2} \label{a}   
\end{equation}
Evidently this term vanishes when $Q\rightarrow 0$.
\subsubsection{Joint terms }
To compute the contribution from the various joints, we need the inner products of the normal to the singularity with the null generators of the null boundary of the WdW:
\begin{align}
    k_+k_-&=\frac{-2l^2}{f(r)}\nonumber\\
    k_+\cdot n &=\frac{l}{f^{1/2}(r)}\nonumber\\
     k_-\cdot n &=\frac{-l}{f^{1/2}(r)}.\nonumber
\end{align}
The contribution from the intersection of the timelike singularity and null boundaries of the WdW patch are then,
\begin{align}
    I_{j\pm}&=\lim_{r\to 0}\frac{1}{8\pi G_N}\int dx\,dyf(r)\log\left|\frac{l}{f^{1/2}(r)}\right|= 0.
\end{align}
The contribution from the intersection of the null boundaries of the WdW patch,
\begin{small}
\begin{align}
    I_{nn}&=-\lim_{r\to \Lambda}\frac{1}{8\pi G_N}\int dx dy f(r)\log\left|\frac{-l^2}{f(r)}\right| \nonumber \\
    &=\frac{\Lambda^2 V_{xy}}{8 \pi  l^2 G_N} \left(\frac{\Lambda}{l}\right)+\frac{\Lambda  V_{{xy}} \left(2Q \ln \left(\frac{\Lambda}{l}\right)+Q\right)}{8 \pi  \left(\alpha ^2+1\right) l^2 G_N}-\frac{V_{xy} \left(2\alpha ^2 Q^2 \ln \left(\frac{\Lambda}{l}\right)-2Q^2 \ln \left(\frac{\Lambda}{l}\right)+(\alpha ^2-3) Q^2\right)}{16 \pi  \left(\alpha ^2+1\right)^2 l^2 G_N} \label{ED NN}
\end{align}
\end{small}
\subsubsection{Full action contribution: Analytical and numerical estimates}

Combining (\ref{ED EH}) (\ref{eq: Scalar kinetic term}), (\ref{eq: Scalar potential term}), \eqref{ED LMPS} and \eqref{ED NN} we obtain the following contribution towards the action complexity of the Einstein Scalar system
\begin{align}\label{ESCA}
  C_{\mathcal{A}}=\frac{V_{xy}}{8\pi^2 G_N }  \left[2 (\ln 2-1) \frac{\Lambda ^2}{l^2}+\frac{4 (\ln2-1) }{  \left(\alpha ^2+1\right) }\frac{\Lambda  Q}{l^2}+\frac{6 \alpha ^2 \epsilon ^{\frac{3 \alpha ^2-1}{\alpha ^2+1}} Q^{\frac{3-\alpha ^2}{\alpha ^2+1}}}{\left(3 \alpha ^2-1\right)}+\frac{2 \alpha ^2 Q^2}{(1+\alpha^2)^2} \ln\left(\frac{\Lambda }{l}\right)+\text{finite} \right]
\end{align}
where the finite part is
\begin{align*}
   \frac{1}{l^2(1+\alpha^2)^2} \Bigg[a \left(\alpha ^2+1\right)^2 l^2+12 \alpha ^2 \left(3 \alpha ^2-1\right) Q^2 \left(\psi ^{(0)}\left(3-\frac{4}{\alpha ^2+1}\right)-\psi ^{(0)}\left(\frac{4 \alpha ^2}{\alpha ^2+1}\right)\right)\\\nonumber+Q^2 \left(11 \alpha ^4+\alpha ^2 (2 \gamma -2 \gamma +17-4 \log (2))+\log (4)\right)\Bigg]\nonumber
\end{align*}
It is immediate from (\ref{ESCA}) that the complexity of the geometry containing the timelike singularity arising from the Einstein Scalar system is positive in the range $\alpha>1/\sqrt{3}$.  Moreover, in the same range of values of $\alpha$, the action complexity is greater than the complexity of the empty AdS spacetime which can be obtained by taking the $Q=0$ in the above expression. And quite interestingly, this is the same range of values where the geometry is in the allowed phase according to the Gubser criterion. In the complementary range, $\alpha<1/\sqrt{3}$ the action complexity criterion adheres to the Gubser criterion by furnishing the value of negative infinity thereby signalling that the phase the geometry is in the disallowed range.

We also supplement our analytical result with the numerical values as listed in the table (\ref{CA ES table})
where it can be confirmed that the action complexity scales as $\sim \Lambda^{2}$ in the range of allowed values according to the Gubser criterion and there is virtually no dependence on $Q$! The numerics also confirm that complexity undergoes a dramatic change
in behavior across the Gubser bound on $\alpha$. The numerical estimates also adhere to our conjectured action complexity criterion as is evident from the numerical values listed in the last column where it can be easily seen that the complexity virtually blows up to the unreasonably large negative values.

\begin{table}[h]
 	\centering
 	\begin{tabular}{|l|c|c|c|r|}
 		\hline
 		{$Q$} &$\alpha$ &  $\Lambda$& $\frac{G_N \hbar}{V_{xy}}C_{A}$&$\frac{G_N \hbar}{V_{xy}} (C_{A}-C_{A}^{AdS})$\\
 		\hline
 		\hline
 		10 &0.1 &$ 10^{6}$&$-1.13\times 10^{230}$&$-1.13\times 10^{230}$\\
 		\hline
 		10 &0.4 & $ 10^{6}$&$-1.17\times 10^{110}$&$-1.17\times 10^{110}$\\
 		\hline
 		10& 0.5& $ 10^{6}$&$-3.32\times 10^{51}$&$-3.32\times 10^{51}$\\
 		\hline
 	10& 0.6& $ 10^{6}$&$2.66\times 10^{13}$&$4.06\times 10^8$\\
 		\hline
 		10 & 0.9&$ 10^{6}$&$2.66\times 10^{13}$&$3.05\times 10^{8}$\\
 		\hline
 		10 &0.1 & $ 10^{10}$&$-1.62\times 10^{226}$&$-1.62\times 10^{226}$\\
 		\hline
   10 &0.4 & $ 10^{10}$&$-1.88\times 10^{108}$&$-1.88\times 10^{108}$\\
 		\hline
 		10&0.5 & $10^{10}$ & $-5.26\times 10^{50}$ & $-5.26\times 10^{50}$\\
 		\hline
 		10& 0.6 & $ 10^{10}$&$4.50\times 10^{21}$&$6.77\times 10^{12}$\\
 		\hline
 		10 & 0.9 &$10^{10}$&$4.50\times 10^{21}$&$5.08\times 10^{12}$\\
 		\hline
 		1000& 0.1 & $10^{6}$&$-9.40\times 10^{235}$&$-9.40\times 10^{235}$\\
 		\hline
 
 		1000 &0.4 &$  10^{6}$&$-9.21\times 10^{114}$&$-9.21\times 10^{114}$\\
 		\hline
 		1000&0.5 & $10^{6}$&$-8.34\times 10^{55}$&$-8.34\times 10^{55}$\\
 		\hline
 		1000&0.6& $10^{6}$& $2.66\times 10^{13}$& $4.06\times10^{10}$\\
 		\hline
 	1000& 0.9& $10^{6}$&$2.66\times 10^{13}$&$3.05\times 10^{10}$\\
 		\hline
 		1000&0.1&$10^{10}$&$-1.35\times 10^{232}$&$-1.35\times 10^{232}$\\
 		\hline
 		1000 & 0.4 & $ 10^{10}$&$-1.48\times 10^{113}$&$-1.48\times 10^{113}$\\
 		\hline
   1000 &0.5& $10^{10}$&$-1.32\times 10^{55}$&$-1.32\times 10^{55}$\\
 		\hline
 		1000& 0.6 & $ 10^{10}$& $4.50\times 10^{21}$&$6.77\times 10^{14}$\\
 		\hline
 		1000& 0.9 & $ 10^{10}$&$4.50\times 10^{21}$&$5.09\times 10^{14}$\\
 		\hline
 	\end{tabular}
 \caption{Table for dependence of action complexity upon $Q,\,\alpha$ and cut off $\Lambda$ for the Einstein-Scalar system.}
 \label{CA ES table}
 \end{table}
 
 Therefore, we see that the action complexity successfully detects as the geometry transitions from the allowed phase to the disallowed phase across the Gubser point of $\alpha=1/\sqrt{3}$. It remains for us to see whether the volume complexity also follows the same trend and registers the agreement with the Gubser criterion by reproducing the same result as the action complexity in the next subsection. 
\subsection{Volume Complexity for the Einstein-Scalar system}
\begin{figure}[h]
\includegraphics[width=0.5\textwidth]{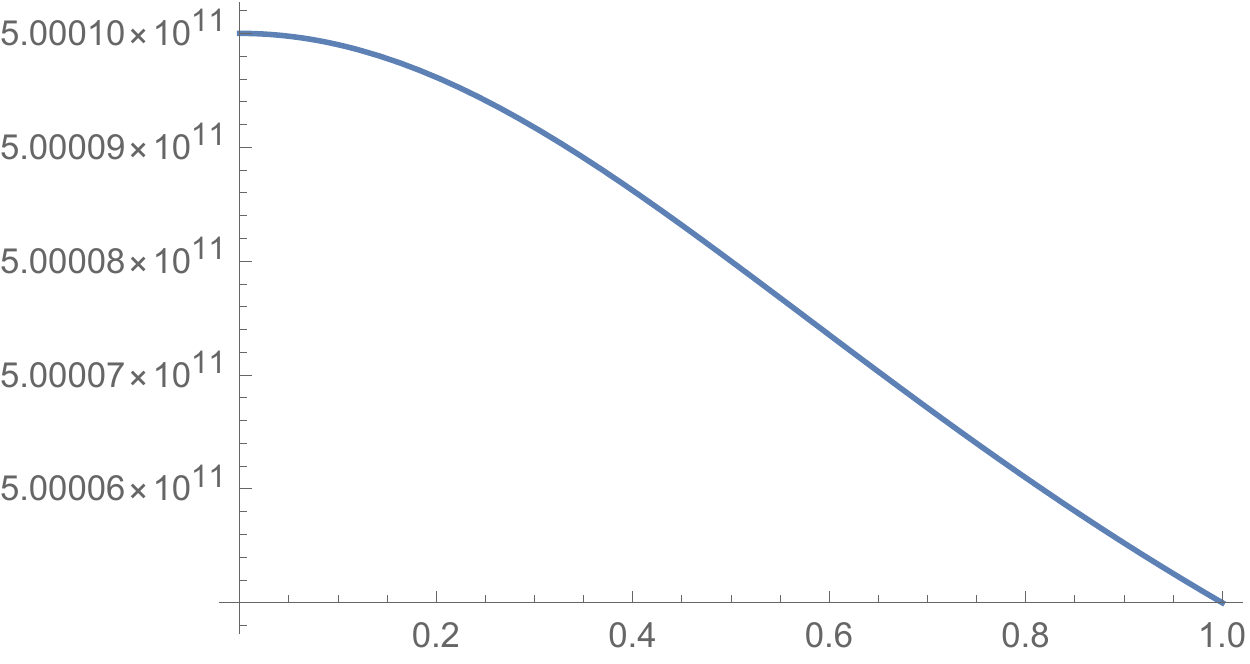}
\includegraphics[width=0.5\textwidth]{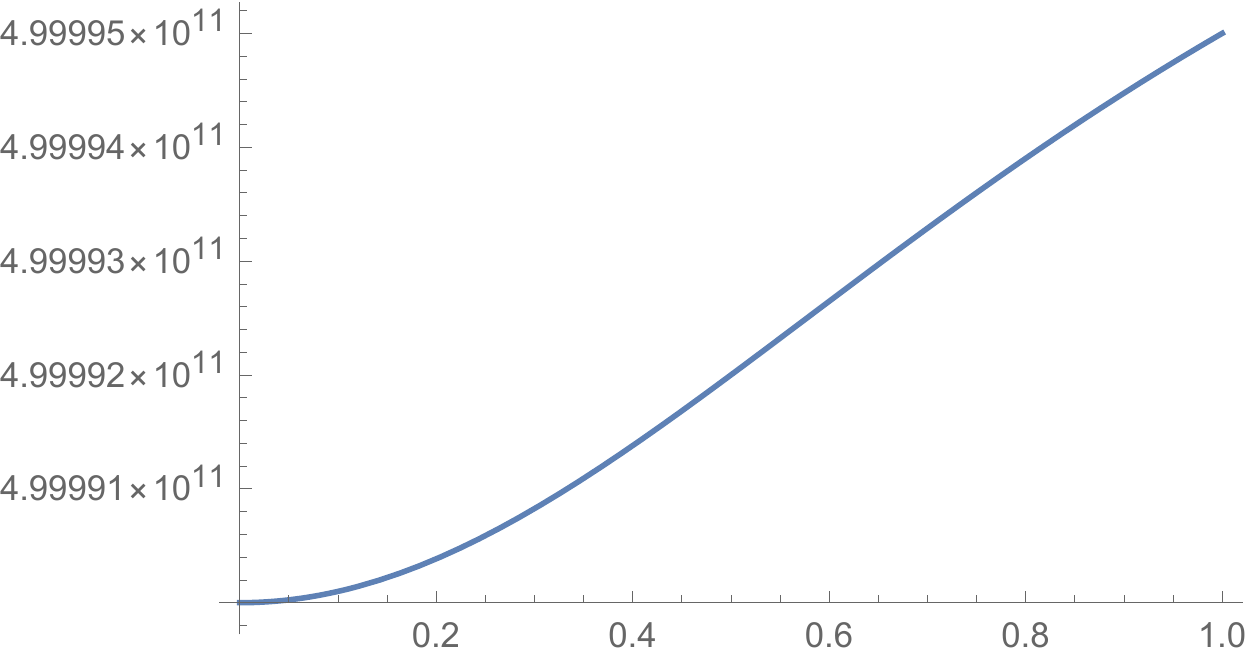}
\includegraphics[width=0.5\textwidth]{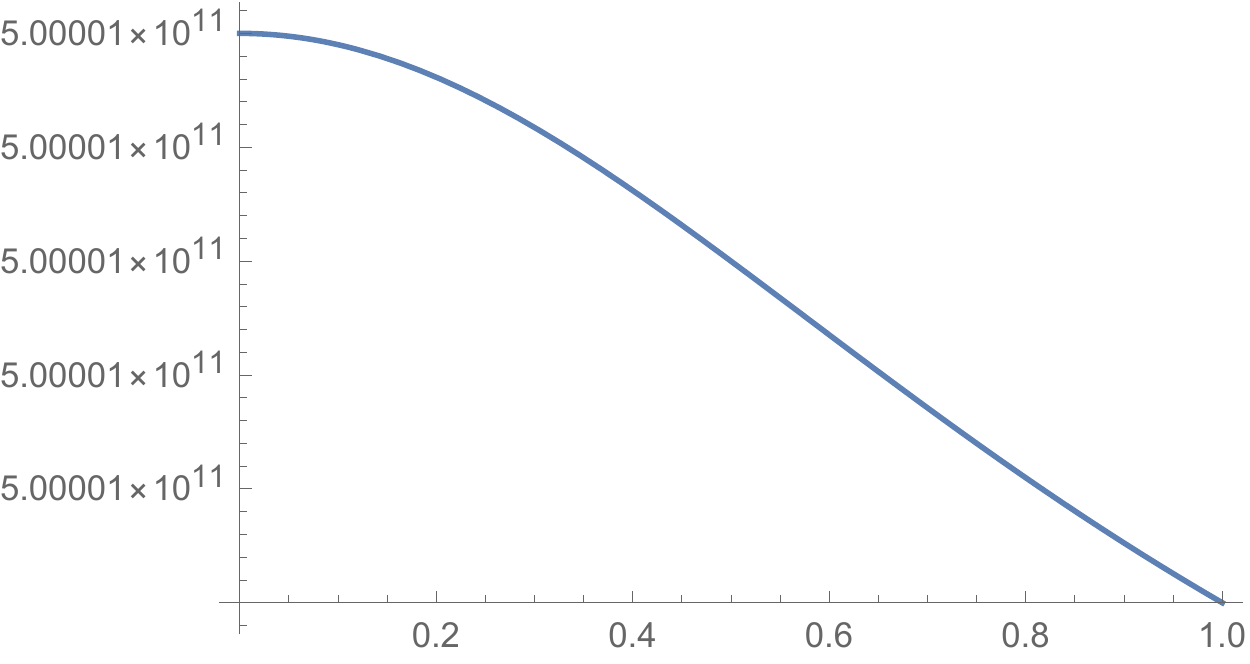}
\includegraphics[width=0.5\textwidth]{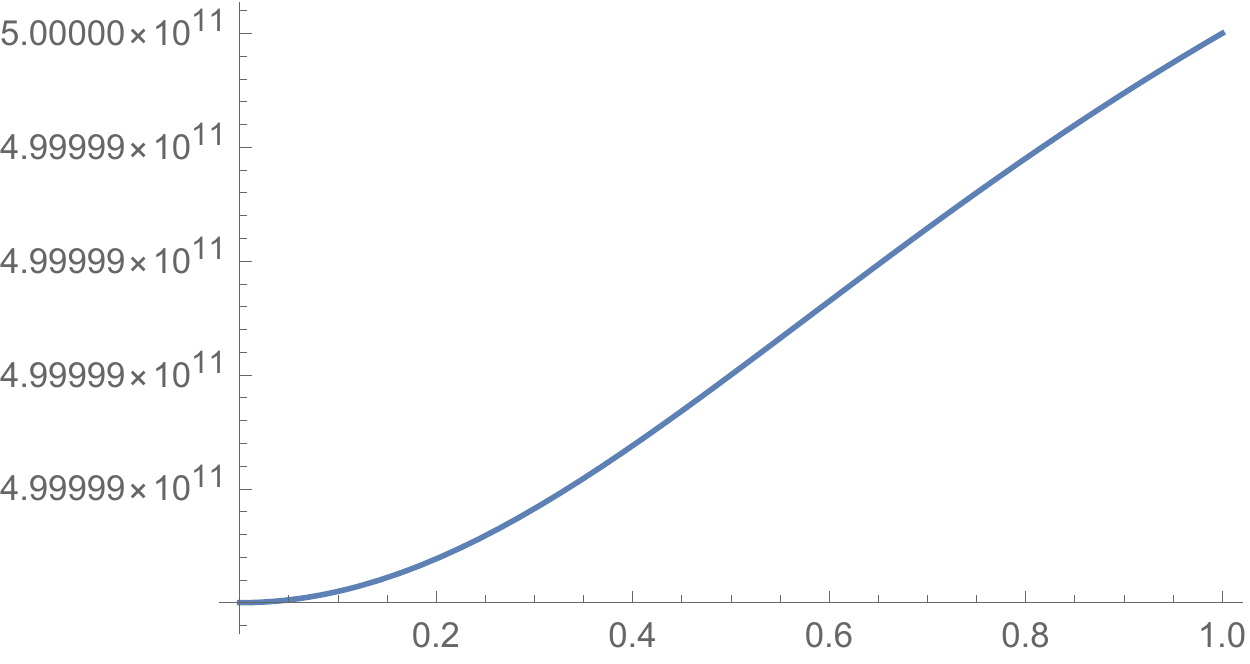}
\includegraphics[width=0.5\textwidth]{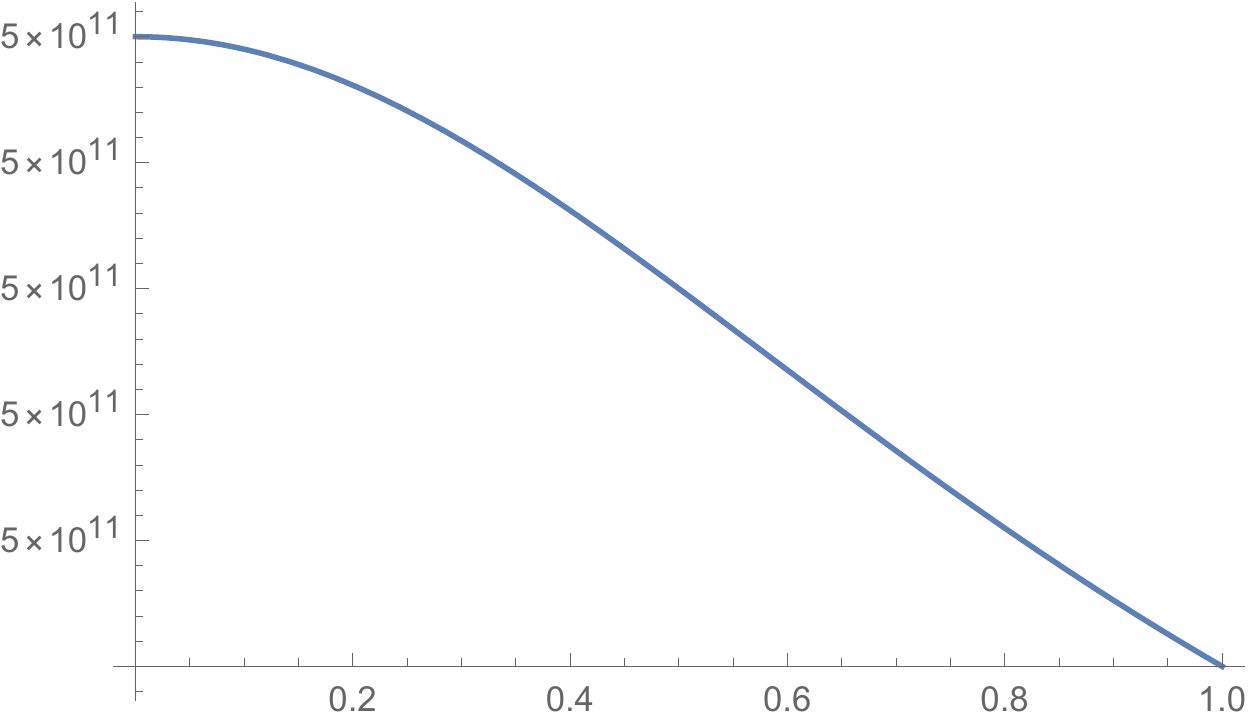}
\includegraphics[width=0.5\textwidth]{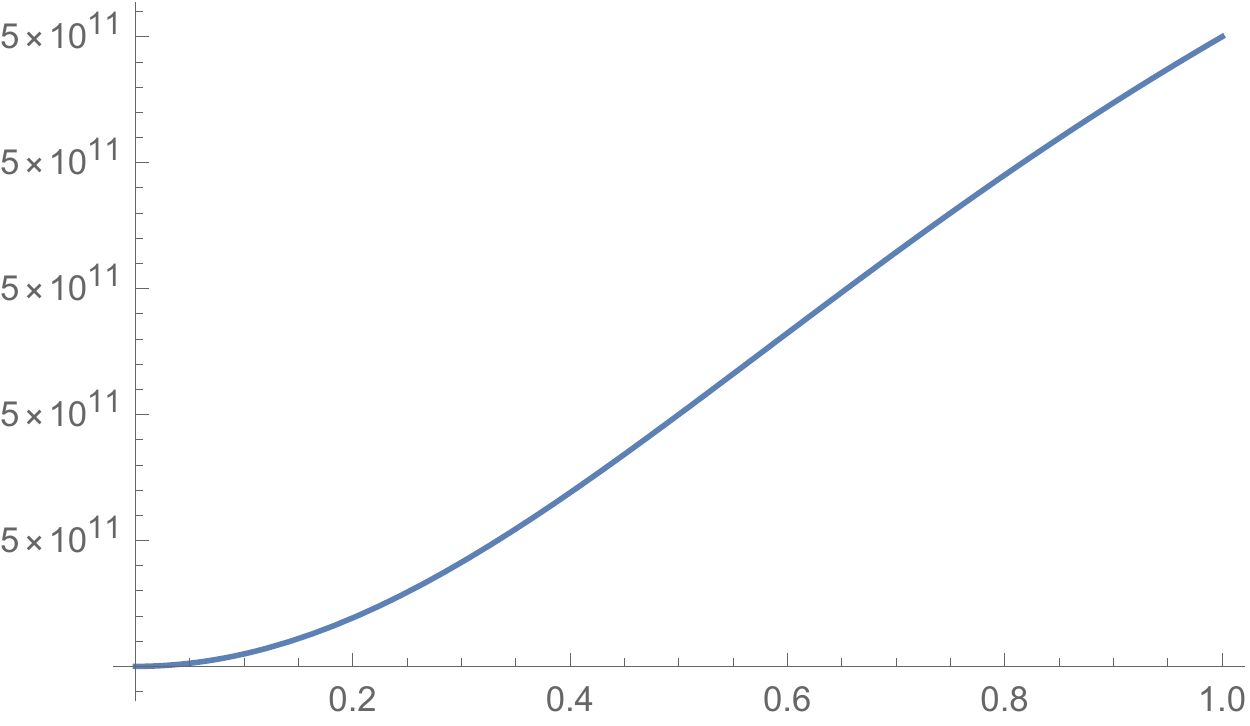}
\caption{Volume (Complexity) plots as a function of the exponent, $\alpha$
for different values of $Q$. The top row is $|Q|=10$, middle row
is $|Q|=1.0$ and the bottom row is $|Q|=0.1$. The left panel are
for $Q>0$, while the right panel is for $Q<0$. \label{fig:Volume-Complexity-plots}}
\end{figure}

It will be interesting to compute the volume complexity to see if
it reproduces the same behavior as action complexity. Consider a codimension one spatial hypersurface specified by the condition $t=t(r),\forall x,y$. The pullback of (\ref{ED_metric}) on this hypersurface is
\begin{align*}
    ds^2=\left(\frac{1}{f(r)}-f(r) t'(r)^2\right)dr^2+f(r)\,d\boldsymbol{x}^2~.
\end{align*}
The volume of the spacelike slice turns out to be
\[
V=V_{xy}\int_{0}^{\infty}dr\;f(r)\sqrt{\frac{1}{f(r)}-f(r)t'^{2}(r)}~.
\]
Evidently the volume is maximized when the   quantity inside the square root is maximized, i.e. $t'(r)=0$. Thus the maximal volume spatial slice is the constant time slice, $t(r)=T$, where, $T$ is the time value at which the spatial slice meets the spatial boundary (where the WDW patch is anchored).\\
Therefore, the maximal volume slice volume is
\[
V_{max}=V_{xy}\int_{0}^{\infty}dr\;\sqrt{f(r)}~.\nonumber
\]
This integral is clearly UV divergent, so we introduce a UV (bulk IR) cutoff, $\Lambda$.
Then,
\begin{equation}
\begin{split}
V_{max} & =\frac{V_{ xy}}{l}\int_{0}^{\Lambda}dr\;r\left(1+\frac{Q}{r}\right)^{\frac{1}{1+\alpha^{2}}}\label{eq: Maximal volume for Einstein-Scalar case}\\
 & =\frac{V_{ xy}}{l}\left[\frac{\Lambda^{2}}{2}+\frac{Q}{1+\alpha^{2}}\Lambda-\frac{\alpha^{2}Q^{2}}{2\left(1+\alpha^{2}\right)^{2}}\ln\Lambda - \frac{\alpha^{2}Q^{2}}{4\left(1+\alpha^{2}\right)^{2}}\left(3-2H_{\frac{\alpha^2}{1+\alpha^2}}\right)+O(\Lambda^{-1})\right].
\end{split}
\end{equation}
The volume complexity is,
\begin{equation}
C_{V}\equiv\frac{V_{max}}{G_{N}\,l}=\frac{V_{xy}}{G_{N}}\left[\frac{\Lambda^{2}}{l^2}+\frac{Q/l}{1+\alpha^{2}}\frac{\Lambda}{l}-\frac{\alpha^{2}Q^{2}/l^2}{2\left(1+\alpha^{2}\right)^{2}}\ln\frac{\Lambda}{l}- \frac{\alpha^{2}Q^{2}/l^2}{4\left(1+\alpha^{2}\right)^{2}}\left(3-2H_{\frac{\alpha^2}{1+\alpha^2}}\right)\right]. \label{VC_ES}
\end{equation}
From this expression we note the following features of volume complexity:
\begin{itemize}
\item The volume complexity has a leading quadratic UV-divergence, $\frac{V_{xy}}{G_{N}}\frac{\Lambda^{2}}{2 l^2}$,
as expected for an extensive quantity in a $2+1$-dimensional field theory dual and \emph{manifestly positive}. There are subleading linear and log divergent pieces and a finite piece.
\item The leading UV-divergent contribution is also \emph{independent of Q} and is identical to a pure AdS volume complexity. The subleading piece is positive definite for positive $Q$, i.e. higher complexity than that of empty AdS.
\item Numerical computations for some fixed large value of the UV
cutoff $\Lambda$, say $\Lambda\sim10^{10}$ in AdS units, and for several
distinct values of $Q>0$, the complexity is a monotonically decreasing
function of $\alpha$, while $Q<0$ the volume complexity is monotonically
increasing function of $\alpha$. The results are displayed in Fig.
\ref{fig:Volume-Complexity-plots}.
\item There is no contribution from the naked timelike singularity at $r=0$ - since the volume crunches at the singularity! This is evident from the volume intgeral, which receives vanishing contribution at the lower limit $r=0$.
\item According to the Gubser criterion, for $\alpha<\frac{1}{\sqrt{3}}$, the naked timelike singularity is not resolvable or allowable, in the sense that it cannot be resolved by embedding in string/M
theory. However, as is evident from the $C_{V}$ expression \eqref{VC_ES} or the $C_{V}$
vs. $\alpha$ plots in Fig. \ref{fig:Volume-Complexity-plots},
the volume complexity does not seem to display any sudden change in pattern
as we vary $\alpha$ across the Gubser bound,
$\alpha=\frac{1}{\sqrt{3}}$ in all cases.  
\end{itemize}
The results of this sections leads us to conclude that volume complexity is not a sensitive probe of timelike singularities, i.e. whether they are resolvable (allowable) in a fundamental quantum theory (UV-completion) of gravity. The action complexity too is mostly unreliable except for this example, where it is able to show the transition from allowable to non-allowable singularities.
 \section{Conclusion \& Outlook} \label{cc}
The complexity has recently been added in the arsenal of holographic tools available to study quantum gravity \cite{Susskind:2014rva,Brown:2015bva,Brown:2015lvg} because of its anticipated role in addressing unresolved issues in Quantum Gravity when viewed in the light of Quantum Information Theory. However, its potential scope in closing crucial gaps in our understanding of quantum gravity compared to other entanglement measures, such as entanglement entropy, is still being investigated and explored. There are reasons to believe that complexity can shed light on the nature of gravitational singularities in general, a topic in which so far holographic approach to quantum gravity has not been particularly fruitful. In this work, have tried to explore holographic complexity as a tool in probing naked timelike singularities, the status of which are not clear in traditional string theory approaches. In past works, the question of the existence of the naked timelike singularities, particularly in holographic contexts, has been attacked with the tool of Gubser criterion \cite{Gubser:2000nd, Gursoy:2008za}, which constitutes a set of diagnostic conditions for allowing or disallowing a particular case of the singular geometries. With the anticipation that the physics of complexity will also supply a testing ground alongside the Gubser criterion, we probed the well-known sick geometry of negative mass Schwarzschild AdS black hole \cite{Horowitz:1995ta} using action complexity and volume complexity as a warm-up example. In its study, we found that both prescriptions of complexity provided us with what would appear as a physically acceptable result (a well-behaved positive complexity with the appropriate divergence structure). On a closer inspection of the volume complexity, one notes that the singular geometry was less complex than the empty AdS background! This led us to speculate about a \textit{complexity criterion}, namely that if the singular geometries have lower complexity compared to the empty AdS backgrounds, then they are not allowed or more accurately, cannot be realized in the classical limit of theory of quantum gravity. Complexity represents a measure of the ease of creating a state (bulk geometry) from some specified reference state. A singular bulk geometry which is less ``complex'' than empty AdS implies that the cost of creating a singular geometry is less than the cost of creating empty AdS space. From our negative mass AdS-Schwarzschild exercise, it was suggestive (but not definitive) that this would be undesirable in a UV-complete theory of (AdS) quantum gravity where the vacuum state, i.e. a semiclassical smooth pure/empty AdS spacetime is stable and will have lowest holographic complexity among all allowable semiclassical geometries (with the caveat of compactifying (extra) spatial dimensions as in \cite{Engelhardt:2021kyp} whereby one can lower the complexity). Armed with this insight, we went ahead in the next section to focus on an anisotropic geometry occurring in the effective holographic theories namely, timelike Kasner AdS spacetime \cite{Ren:2016xhb}. According to Gubser criterion, this geometry is not a allowable singularity. Here, just as was in the case of negative mass SAdS geomtery, the action complexity appears perfectly physical - revealing no sign of pathology anywhere (notably the contribution to the complexity from the singularity was finite but negative). The volume complexity of the timelike Kasner also at first glance appears perfectly fine. However, just like in the negative mass SAdS case, in a restricted range of the Kasner exponent, namely $\alpha<2/3$ is less than that of pure (Poincar\'e) AdS thereby potentially signalling a pathology. In the hopes to settle this issue, in the next section, we explored asymptotically AdS timelike singularities which arise in the Einstein-Scalar system  \cite{Gao:2004tu,Ren:2019lgw} in holographic condensed matter studies. Such a smooth bulk background have been extensively investigated in the light of the Gubser criterion and is known to exhibit the transition from the disallowed to the allowed geometry as the relevant parameter, $\alpha$ is tuned. Here the action-complexity too undergoes a sharp transition across the Gubser point. In the allowable range, complexity is positive and receives finite contribution from the singularity while in the disallowed range complexity received a negative and divergent contribution from the timelike singularity. This final case study of the Einstein-Scalar system suggests a striking agreement with the Gubser criterion: the action complexity criterion is successfully able to identify the Gubser point. On the other hand, the volume complexity is not able to register the transition of the geometry hinting that the volume complexity as a probe is not very suitable for investigation of the timelike naked singularities.

Based on the evidence gained from studying the various cases, we infer that neither the volume-complexity nor action-complexity can furnish a reliable probe in the investigation of the sick geometries alongside the Gubser criterion. However just as in the Einstein-Dilaton system, i.e. for isolated cases the action-complexity can diagnose pathological (inadmissible) timelike naked singulaities by registering negative divergent contributions from the singularity. This action complexity criterion is certainly operationally much easier to implement than Gubser criterion itself, since given a spacetime containing a (naked) timelike singularity, it is not immediately obvious whether that geometry can or cannot be realized as the extremal limit of some black-hole geometry - one just needs to compute the onshell action supported in the WdW patch and compare to the pure AdS! However in this work we have only scratched the surface as far as timelike singularities and their holographic complexity features are concerned, looking at three simple examples. We have to conduct a more comprehensive survey of various other geometries with timelike singularities, beyond even the asymptotically AdS examples to check the correlation of the allowable singularities (as per Gubser) with negative divergent contribution to action-complexity from the timelike naked singularity. We leave such an exhaustive study for future work(s). Even so, the bottom line is that neither action-complexity nor volume-complexity can unambiguously diagnose timelike singularities which can be allowed in a UV complete theory of QG.

 \section*{Acknowledgements}

The work of GK was supported partly by a Senior Research Fellowship (SRF) from the Ministry of Education (MoE), Govt. of India and partly from the RDF fund of SR: RDF/IITH/F171/SR. The work of SR is supported by the IIT Hyderabad seed grant SG/IITH/F171/2016-17/SG-47. The work of JR is supported partly by the NSF of China under Grant No. 11905298. We thank Aasmund Folkestad for bringing the work \cite{Engelhardt:2021kyp} to our attention. We also thank the anonymous referee for his/her perceptive remarks and suggestions which led to significant improvements to the draft.

 \appendix
\section{Perturbative analysis for negative mass SAdS complexity}\label{App.A}

Here we perform an analytic computation of the action and volume complexity for the negative mass AdS-Schwarzschild black hole by performing a perturbation to first two orders in the parameter $\mu/\Lambda^{D-3}$. Our aim is to confirm, at least to leading order in perturbation theory, the numerical results that the negative mass Schwarzschild-AdS geometry has a lower complexity than empty AdS.\\\\
\textbf{Action Complexity}: The Einstein-Hilbert term \eqref{SAdS-EH} works out to be,
\begin{equation*}
I_{EH} = I_{EH}^0+\mu \Lambda^3\frac{\left(D-1\right)\Omega_{D-2}}{4\,\pi\,G_N\,l^2} \int_0^1 dy \,y^{D-2} \int_y^1\frac{dz}{z^{D-3}(1+(z\Lambda/l)^2)^2}.
\end{equation*}
The nested integral can be evaluated for arbitrary $D$,
\begin{equation}
I_{EH} = I_{EH}^0+\frac{\Omega_{D-2}}{16\,\pi\,G_N} \pi\, \mu\, l -\frac{\Omega_{D-2}}{4\,\pi\,G_N} \frac{\mu\, l}{\Lambda}+O(l^2/\Lambda^2).
\end{equation}
Thus this EH term complexity contribution is larger than the empty AdS EH term complexity contribution.
The GHY-term arising from the null boundaries of the WdW patch \eqref{SAdS-GHY}, to first two orders, works out to be
\begin{equation*}
 I_{GHY}^{\partial WdW}= I_{GHY}^{0\,\partial WdW}-\frac{\Omega_{D-2}\,\mu\,\Lambda}{16\pi G_{N}}\int_{0}^{1}dy\: \frac{\left((D-3) +(D-1) y^2/a^2\right)}{\left(1+y^2/a^2\right)^2}
\end{equation*}
where $a=l/\Lambda$. This integral can be computed exactly for arbitrary $D$ and we get,
\begin{equation}
I_{GHY}^{\partial WdW}= I_{GHY}^{0\,\partial WdW}-\frac{\Omega_{D-2}}{16\pi G_{N}} \frac{(D-2)\,\pi\, \mu\,l}{2}+\frac{\Omega_{D-2}}{16\pi G_{N}} \frac{(D-1)\, \mu\,l}{\Lambda}+O(l^2/\Lambda^2)
\end{equation}
Thus to subsubleading order in $\mu/\Lambda^{D-3}$, the action-complexity for the negative mass Schwarzschild-AdS geometry is is evidently lower than empty AdS complexity,
\begin{equation}
\mathcal{C}_A = \mathcal{C}_A^{AdS}-\frac{\Omega_{D-2}}{16\pi G_{N}} \frac{(D-4)\, \mu\,l}{2} +\frac{(D-5)\,\Omega_{D-2}}{16\pi G_{N}} \frac{\mu\,l^2}{\Lambda} +O(l^2/\Lambda^2).
\end{equation}
Note that $\mathcal{C}_A^{AdS}\sim O\left((\Lambda/l)^{D-2}\right)$, so the linear order term in $\mu$ is suppressed by a factor of $\mu\,l/\Lambda^{D-2}$. For the special case of $D=4$, the linear (leading) order in $l/\Lambda$ difference vanishes, however the negative mass SAdS still receives a negative contribution from the subleading term and has a lower complexity than empty AdS$_4$.\\\\
\textbf{Volume complexity}: The volume complexity expression \eqref{CV nmSAdS} to linear order in $\mu$ is,
\begin{equation*}
\mathcal{C}_V=\mathcal{C}_V^0 -\frac{\mu\,\Omega_{D-2}}{2\,G_N\,l}\int^{\Lambda}_{0}dr\, \frac{r}{\left(1+\frac{r^2}{l^2}\right)^{3/2}}=\mathcal{C}_V^0 (T)-\frac{\Omega_{D-2}\,\mu\,l}{2\,G_N}+O\left(l^2/\Lambda^2\right).
\end{equation*}
Evidently this is lesser that empty AdS complexity, $\mathcal{C}^0_V \sim \Lambda^2$.

\end{document}